\documentclass[twocolumn]{aastex62}

\renewcommand\edit{}

\graphicspath{{./}{figures/}}

\received{September 4, 2019}
\revised{October 18, 2019}
\accepted{November 7, 2019}
\submitjournal{ApJ}

\shorttitle{Irregular structures around pre-main-sequence stars}
\shortauthors{Laws et al.}

\begin{document}

\title{Irregular dust features around intermediate-mass young stars with GPI: signs of youth or misaligned disks?}

\correspondingauthor{Anna S.E. Laws}
\email{al630@exeter.ac.uk}

\author[0000-0002-2145-0487]{Anna S.E. Laws}
\affil{Astrophysics Group, University of Exeter, Stocker Road, Exeter, EX4 4QL, U.K.}

\author[0000-0001-8228-9503]{Tim J. Harries}
\affiliation{Astrophysics Group, University of Exeter, Stocker Road, Exeter, EX4 4QL, U.K.}

\author[0000-0001-5980-0246]{Benjamin R. Setterholm}
\affiliation{Department of Astronomy, University of Michigan, West Hall, 1085 South University Ave, Ann Arbor, MI 48109-1090, U.S.A.}

\author[0000-0002-3380-3307]{John D. Monnier}
\affiliation{Department of Astronomy, University of Michigan, West Hall, 1085 South University Ave, Ann Arbor, MI 48109-1090, U.S.A.}

\author[0000-0002-1779-8181]{Evan A. Rich}
\affiliation{Department of Astronomy, University of Michigan, West Hall, 1085 South University Ave, Ann Arbor, MI 48109-1090, U.S.A.}

\author[0000-0002-1327-9659]{Alicia N. Aarnio}
\affiliation{University of North Carolina Greensboro, Greensboro, NC, 27402, U.S.A.}

\author[0000-0002-8167-1767]{Fred C. Adams}
\affiliation{Department of Astronomy, University of Michigan, West Hall, 1085 South University Ave, Ann Arbor, MI 48109-1090, U.S.A.}
\affiliation{Physics Department, University of Michigan, Ann Arbor, MI 48109, U.S.A}

\author[0000-0003-2253-2270]{Sean Andrews}
\affiliation{Center for Astrophysics \textbar Harvard \& Smithsonian, 60 Garden Street, Cambridge, MA 02138, U.S.A.}

\author[0000-0001-7258-770X]{Jaehan Bae}
\affiliation{Department of Terrestrial Magnetism, Carnegie Institution for Science, 5241 Broad Branch Road, NW, Washington, DC 20015, U.S.A.}

\author[0000-0002-3950-5386]{Nuria Calvet}
\affiliation{Department of Astronomy, University of Michigan, West Hall, 1085 South University Ave, Ann Arbor, MI 48109-1090, U.S.A.}

\author[0000-0001-9227-5949]{Catherine Espaillat}
\affiliation{Boston University, Boston, MA, U.S.A.}

\author[0000-0003-1430-8519]{Lee Hartmann}
\affiliation{Department of Astronomy, University of Michigan, West Hall, 1085 South University Ave, Ann Arbor, MI 48109-1090, U.S.A.}

\author[0000-0001-8074-2562]{Sasha Hinkley}
\affiliation{Astrophysics Group, University of Exeter, Stocker Road, Exeter, EX4 4QL, U.K.}

\author[0000-0001-8061-2207]{Andrea Isella}
\affiliation{Department of Physics \& Astronomy, Rice University, 6100 Main Street, Houston, TX, 77005, U.S.A.}

\author[0000-0001-6017-8773]{Stefan Kraus}
\affiliation{Astrophysics Group, University of Exeter, Stocker Road, Exeter, EX4 4QL, U.K.}

\author[0000-0003-1526-7587]{David Wilner}
\affiliation{Center for Astrophysics \textbar Harvard \& Smithsonian, 60 Garden Street, Cambridge, MA 02138, U.S.A.}

\author[0000-0003-3616-6822]{Zhaohuan Zhu}
\affiliation{University of Nevada, Las Vegas, NV, U.S.A.}



\begin{abstract}


We are undertaking a large survey of over thirty disks using the Gemini Planet Imager (GPI) to see whether the observed dust structures match spectral energy distribution (SED) predictions and have any correlation with stellar properties.
GPI can observe near-infrared light scattered from dust in circumstellar environments
using high-resolution Polarimetric Differential Imaging (PDI) with coronagraphy and adaptive optics.
The data have been taken in \textit{J} and \textit{H} bands over two years,
with inner working angles of 0.08\,'' and 0.11\,'' respectively. 
Ahead of the release of the complete survey results, here we present five objects with extended and irregular dust structures
within 2\,'' of the central star.
These objects are: \object{FU~Ori}; \object{MWC~789}; \object{HD~45677}; \object{Hen~3-365}; and \object{HD~139614}.
The observed structures are consistent with each object being a pre-main-sequence star with protoplanetary dust.
The five objects' circumstellar environments could result from extreme youth and complex initial conditions, from asymmetric scattering patterns due to shadows cast by misaligned disks,
or in some cases from interactions with companions.
We see complex $U_\phi$ structures in most objects that could indicate multiple scattering or result from the illumination of companions.
Specific key findings include
the first high-contrast \edit{observation} of MWC~789 revealing a newly-discovered companion candidate and arc,
and two faint companion candidates around Hen~3-365.
These two objects should be observed further to confirm whether the companion candidates are co-moving.
Further observations and modeling are required to determine the causes of the structures.

\end{abstract}

\keywords{
infrared: planetary systems --- 
planetary systems ---
planet-disk interactions --- 
protoplanetary disks ---
techniques: high angular resolution ---
techniques: polarimetric
}


\section{Introduction}

\begin{figure}
	\centering
	\includegraphics[width=0.45\textwidth]{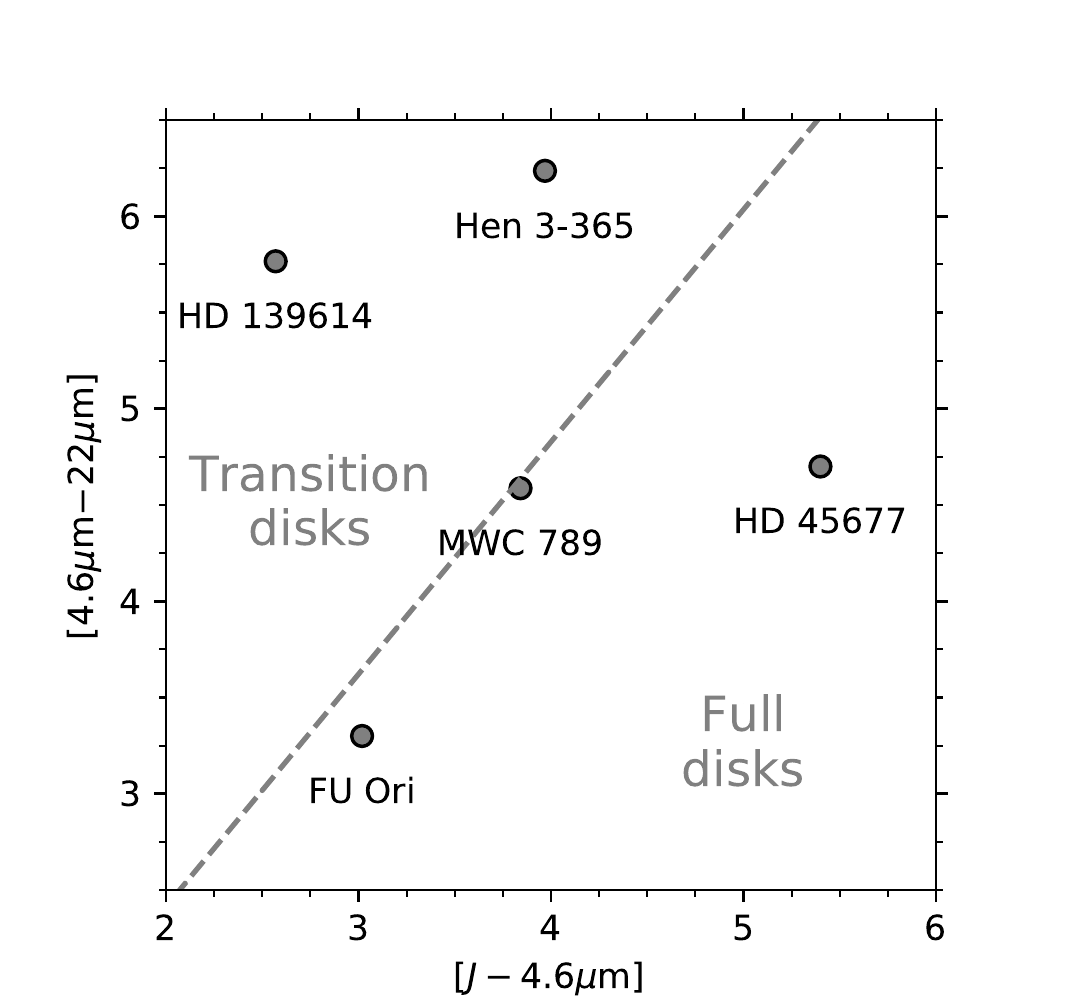}
    \caption{
        A color-color diagram of our five targets to compare the mid-infrared excesses [$4.6\mu$m$-22\mu$m] with the NIR excesses [\textit{J}$-4.6\mu$m].
        The \textit{J}-band, 4.6$\mu$m (WISE2) and 22$\mu$m (WISE4) magnitudes are given in Table~\ref{tab:mags}. 
        The dashed line marks the boundary between predicted full disk and transition disk structures,
        and the slope corresponds to a power law SED for a flat spectrum.
        MWC~789 barely lies on the ``
        full
        disk" side of the boundary.
    }
    \label{fig:colourcolour}
\end{figure}

Protoplanetary disks containing gas and dust surround pre-main-sequence stars as they evolve towards the main sequence \citep[see e.g. review by][]{2011ARA&A..49...67W}.
These disks are the birthplaces of planets, and the large-scale structures we see in the dust \edit{could} provide evidence of where and how planets are forming \citep[e.g.][]{2008ApJ...682L.125E,2011ApJ...729...47Z}.
%
\edit{The structures so far observed include}
 annuli \citep[e.g.][]{2017ApJ...837..132V}, spiral arms \citep[e.g.][]{2013A&A...560A.105G,2015A&A...578L...6B}, arcs \citep[e.g.][]{2016SciA....2E0875L}, and offset rings \citep[e.g.][]{2016A&A...595A.114D,2017ApJ...838...20M}. 
It is then a question of linking these disk structures to the 
\edit{processes that created them.
As well as planet formation,}
other processes
\citep[gravitational torques by stellar mass companions, or self-gravity e.g.][]{1989ApJ...345..554B,2003MNRAS.339.1025R} may also sculpt the disk in the absence of planets.

Disks can be grouped according to their general large-scale structures including large gaps and holes.
In a full state, the dust extends 
broadly
uninterrupted between the pre-main-sequence star and the disk outer edge.
During a pre-transitional stage a large dust-depleted gap 
appears
between the outer and inner parts of the disk. 
Once the inner disk also becomes depleted of dust the system is referred to as a  transition disk,  with a large dust-depleted hole between the central star and an outer dust ring \citep[see review by][]{2014prpl.conf..497E}.
One theory is that disks evolve from a full disk structure through a pre-transitional stage then finally to the transition disk stage, a
process thought to take place over a few million years, with a higher fraction of transition disks appearing in older star-forming regions \citep{2010ApJ...708.1107M}.
To avoid mentioning this evolutionary link, disks are frequently referred to by more general and less weighted terms such as gapped disks and disks with cavities.
Additionally, Herbig Ae/Be stars are classified into Group I and Group II sources based on their large-scale geometry \edit{including their pre/transitional classification} \citep{2001A&A...365..476M}.

It is possible to predict whether a dust disk is full or transitional from the relative strengths of near- and mid-infrared excesses in the object's Spectral Energy Distribution (SED),
as the absence of dust near the star would result in a reduced near-infrared (NIR) excess \citep{1989AJ.....97.1451S,1990AJ.....99.1187S}.
However SEDs only indicate the largest gaps in disks and leave small cavities and details unseen \citep[e.g.][]{2010ApJ...725.1735I},
such as the small gaps and substructures seen by the DSHARP survey \citep[e.g.][]{2018ApJ...869L..41A,2018ApJ...869L..42H}.
So to confirm this prediction 
and detect complicated smaller-scale structures beyond this simplified regime, 
the disks must be observed with direct imaging.



Dust in the upper layers of a disk is visible in infrared light as it scatters light from its central star towards the observer. The scattering events imprint a polarization signature on the unpolarized protostellar light, with a position angle that is perpendicular to the scattering plane, resulting in a centro-symmetric pattern.   Polarimetric Differential Imaging (PDI) may then be used 
to observe faint smaller-scale details in the circumstellar environment that would otherwise be drowned out by the unpolarized starlight
\citep[e.g.][]{2016A&A...595A.113S}.
Typical structures seen with PDI include ring patterns \citep[e.g. TW Hya, ][]{2017ApJ...837..132V} and two distinct spiral arms \citep[e.g. MWC~758, ][]{2015A&A...578L...6B} \citep[see overview by][]{2017Msngr.169...32G}.
Ring structures are seen around lower-mass stars such as T Tauri stars  \citep{2018ApJ...863...44A}.
There have also been instances of objects with multiple spiral arms (\citealp[e.g. HD~142527,][]{2017AJ....154...33A}; \citealp[HD~34700,][]{2019ApJ...872..122M}).
A recent analysis of a large number of scattered light observations from the literature
concludes that the type of disk structures observed depends on the age of the object
(e.g. that spiral arms appear around older objects, 
\citealp{2018A&A...620A..94G}).

\edit{
To find out more about how the structures in disks relate to the underlying processes that sculpt them,
including determining which process is dominant, 
it is necessary to observe disks with a variety of structures and stellar properties.
Observations can reveal evidence of features including misaligned disks, stellar and planetary companions, and extended arms.
In this way it will become possible to build up a complete picture of disk properties and their evolution.
}

\subsection{\edit{Gemini-LIGHTS}}

\edit{The presented work is part of Gemini-LIGHTS (Gemini Large Imaging with Gpi Herbig/T-tauri Survey),}
a systematic survey of intermediate-mass pre-main-sequence stars that are thought to host disks.
We observe using the Gemini Planet Imager (GPI), a NIR instrument on the Gemini South telescope \citep{2014PNAS..11112661M,2015ApJ...799..182P}.
Like most instruments of its kind, GPI uses PDI with extreme adaptive optics to counteract wavefront distortion from atmospheric turbulence and obtain images with resolution close to the diffraction limit \citep{2014PNAS..11112661M}.
%
%
%
%
%
The survey aims to observe a representative sample of over thirty young objects 
to find the frequency of scattered light features,
to link observed features with the objects' SED structures,  
and to search for trends in the disk properties with the stellar masses and cluster ages.


The sample was selected from the full list of $\sim$500 objects from the nearest stellar associations and other T Tauri stars, Herbig Ae/Be, YSOs, and pre-main-sequence stars from the literature.
We selected our final survey targets from this list by considering observational constraints.
The survey targets are nearby young objects with strong NIR and MIR emission 
(i.e. debris disks are excluded) 
and with \edit{ \textit{I}-band magnitude brighter than 9\,mag so that GPI's wavefront sensor can function well}. 
Originally we selected objects with a range of disk types - a roughly equal number of transition disks, pre-transition disks, and full disks.
The full SED shapes and features provide evidence of disk structures out to several 10s of au, 
and in selecting our sample we simplified this by using a color-color diagram to probe disk structures within a few au.
The targets were grouped using the color-color diagram according to the ratio of WISE2-WISE4 to \textit{J}-WISE2 magnitudes, 
where the dividing line between full and transition structures correponds to a power law SED for a flat spectrum in $\lambda\,F_\lambda$ space.
The objects were selected for having significant excesses in the \textit{J}-WISE2 and WISE2-WISE4 color-color plane, and so likely harboring a disk. 
For the presented objects, these magnitudes are given in Table~\ref{tab:mags} and a color-color diagram shown in Figure~\ref{fig:colourcolour}.

 During the course of the \edit{Gemini-LIGHTS} survey we published data on resolved ring structures and  other scattered light detections from Herbig Ae/Be stars \citep{2017ApJ...838...20M},
and recently we have also presented spectacular structures in the disk around HD~34700 including a bright ring with many spiral arms \citep{2019ApJ...872..122M}.
The observed structures typically span several hundred au.

Here we present five objects with significant scattered light, four of which are new detections.  They represent some of the more spectacular examples from our sample due to their especially bright and extended features, and we present an in-depth analysis of each. Note that the observations for the survey are now complete and it will conclude with a statistical analysis of all of the targets in a forthcoming paper.

In Section~\ref{sec:ourtargets} we review the literature on the five  objects, 
followed by Section~\ref{sec:observations} with details of our observations and the data reduction process.
Then in Section~\ref{sec:results} we present results including images of our reduced data, 
with focussed analysis of each object 
and discussion of links between the observed scattered light features and literature predictions. We give our conclusions in Section~\ref{sec:conclusions}.

\section{Our targets} 
\label{sec:ourtargets}

	
\begin{table*}
    \centering
	\caption{Magnitudes of our objects.}
	\label{tab:mags}
	\begin{tabular}{lllll}
		\hline
		Object & \textit{J} & \textit{H} & WISE2 & WISE4 \\
		\hline
        \object{FU~Ori}    & $6.519\pm0.023$ & $5.699\pm0.033$ & $3.509\pm0.065$ & $\phantom{-}0.175\pm0.021$ \\
        \object{MWC~789}   & $8.475\pm0.020$ & $7.528\pm0.026$ & $4.633\pm0.039$ & $\phantom{-}0.046\pm0.015$	 \\
		\object{HD~45677}  & $7.242\pm0.026$ & $6.347\pm0.023$  & $1.812\pm0.030$ & $-2.895\pm0.001$ \\
		\object{Hen~3-365} & $6.217\pm0.037$ & $4.756\pm0.268$ & $2.248\pm0.018$ & $-3.988\pm0.001$  \\
		\object{HD~139614} & $7.669\pm0.026$ & $7.333\pm0.040$ & $5.099\pm0.030$ & $-0.667\pm0.008$  \\
		\hline
	\end{tabular}
    {\par Notes: 
    2MASS \textit{J} and \textit{H}-band magnitudes \citep{2003yCat.2246....0C,2006AJ....131.1163S}, and WISE 2 and 4 magnitudes \citep{2012yCat.2311....0C}.
    }
\end{table*}

\begin{table*}
	\centering
	\caption{
    Background information on our objects.
    }
	\label{tab:background}
	\begin{tabular}{lllll}
		\hline
		Object & Alternate names & RA (J2000) & Dec (J2000) & Spectral type \\
		\hline
        \object{FU~Ori}    & \object{IRAS~05426+0903}, \object{PDS~122}         
        & 05:45:22.4 & $+$09:04:12.3 & F0$^{1}$ \\
        \object{MWC~789}   & \object{IRAS~05591+1630}, \object{HD~250550} 
        & 06:02:00.0 & $+$16:30:56.7 & B9$^{2}$ \\
		\object{HD~45677}  & \object{IRAS~06259$-$1301}, \object{FS~CMa}   
		& 06:28:17.4 & $-$13:03:11.1 & B2$^{3}$ \\
		\object{Hen~3-365} & \object{IRAS~10028$-$5825}, \object{HD~87643} 
		& 10:04:30.3 & $-$58:39:52.1 & B2$^{4}$ \\
		\object{HD~139614} & \object{IRAS~15373$-$4220}, \object{PDS~395}         
		& 15:40:46.4 & $-$42:29:53.5 & A7$^{5}$  \\
		\hline
	\end{tabular}
	{\par Notes: RA (right ascension) and Dec (declination), and spectral type. 
    Spectral type sources are as follows:
    $^{1}$\citet{1977ApJ...217..693H}; $^{2}$\citet{2004AJ....127.1682H};
    $^{3}$\citet{2003ApJS..147..305V};  $^{4}$\citet{1998MNRAS.300..170O}; $^{5}$\citet{1997MNRAS.286..604D}.}
\end{table*}

\begin{table*}
	\centering
	\caption{
	Derived physical parameters of our objects.
    }
	\label{tab:masses}
	\begin{tabular}{llllll}
		\hline
		Object & Distance (pc) & $T_{\textnormal{eff}}$ (K) & Mass (M$_{\sun}$) & Age (Myr) & log(Luminosity) (L$_{\sun}$) \\
		\hline
        FU~Ori    
        & $416.2 ^{+8.6}_{-8.6}$        
        & $4200^{+324}_{-324}$ 
        & $\sim0.3$ 
        & $\sim2$ & 2--2.6 
        \\[5pt]
        MWC~789 
        & $697^{+94}_{-64}$ 
        & $11000^{+500}_{-500}$ 
        & $2.60^{+0.30}_{-0.14}$ 
        & $2.56^{+0.43}_{-0.67}$ 
        & $1.94^{+0.17}_{-0.12}$
        \\[5pt]
		HD~45677  
		& $620^{+41}_{-33}$       
		& $16500^{+3000}_{-80}$ 
		& $4.7^{+1.2}_{-0.4}$ 
		& $0.6^{+3.8}_{-0.3}$ 
		& $2.88^{+0.32}_{-0.17}$ 
		\\[5pt]
		Hen~3-365 
		& $2010^{+570}_{-350}$   
		& $19500^{+5000}_{-3000}$ 
		& $18^{+11}_{-7}$ 
		& $0.020^{+0.052}_{-0.010}$ 
		& $4.60^{+0.64}_{-0.53}$
		\\[5pt]
		HD~139614 
		& $134.7^{+1.6}_{-1.6}$        
		& $7750^{+250}_{-250}$ 
		& $1.481^{+0.074}_{-0.074}$ 
		& $14.5^{+1.4}_{-3.6}$ 
		& $0.773^{+0.032}_{-0.010}$
		\\[5pt]
        %
		\hline
	\end{tabular}
	{\par Notes:
    All values for HD~139614, Hen~3-365, HD~45677 and MWC~789 are from \citet{2018A&A...620A.128V}, where GAIA DR2-derived distances and effective temperature $T_\textnormal{eff}$ have been used in placing the objects on pre-main-sequence stellar tracks to estimate the masses and ages. 
    The exception is FU~Ori with values adopted from the following sources: distance, $T_\textnormal{eff}$, and typical $T_\textnormal{eff}$ uncertainty from \citet{2016A&A...595A...1G,2018A&A...616A...1G}; mass from \citet{2007ApJ...669..483Z}; age from \citet{2012AJ....143...55B}; luminosity from \citet{2018ApJ...864...20T}.
	}
\end{table*}

We present basic data on each of the objects including the distances and infrared magnitudes in Table~\ref{tab:background}, 
while derived physical parameters of the pre-main-sequence stars are given in Table~\ref{tab:masses}.
In the following subsections we provide an overview of the literature on each object individually.

\subsection{FU~Ori}




FU~Ori is the prototype of the FU Orionis class of objects that are characterised by variability 
as they undergo a phase of extremely high mass accretion, predicted to last decades to centuries
\citep{1966VA......8..109H,1977ApJ...217..693H,1985ApJ...299..462H} 
\citep[also see reviews by][]{1996ARA&A..34..207H,2014prpl.conf..387A}. 
%
The distance to the object is 416.2$\pm$8.6\,pc \citep{2016A&A...595A...1G,2018A&A...616A...1G} and
it is composed of a 2\,Myr-old visual binary system where the masses of FU~Ori and its companion FU~Ori~S are 0.3\,M$_\sun$ and 1.2\,M$_\sun$ respectively \citep{2004ApJ...601L..83W,2007ApJ...669..483Z,2012AJ....143...55B}. 
\edit{The two components have a separation of 0.50'' \citep{2004ApJ...601L..83W}.}
Comparing infrared excesses (see Table~\ref{tab:background}, Figure~\ref{fig:colourcolour}) indicates that there is a full disk structure.
The object has an inclination of 52--58$^\circ$ \citep{2005A&A...437..627M,2006ApJ...648..472Q}.

Subaru-HiCIAO scattered light observations of FU~Ori in the \textit{H}-band reveal an asymmetrical  structure 
spanning
$\sim$400\,au with a large-scale stream and a bright north-eastern arm extending $\sim$200\,au, features that are likely a stream or filament in the envelope.
\citep{2016SciA....2E0875L,2018ApJ...864...20T}.
\edit{\citet{2016SciA....2E0875L} conclude} that the features may be caused by gravitational instability,
and that the companion star FU~Ori~S is also seen via scattering in its own circumstellar disk. 
\textit{K}-band observations from the Keck Interferometer reveal that the asymmetric features continue down to the inner 1\,au scale \citep[]{2011ApJ...738....9E}.



FU~Ori has also been the subject of resolved sub-mm and radio observations. 
No substantial circumbinary ring was detected in ALMA band 7 continuum observations, however each star could have its own unresolved disk of maximum radius 0.1\," or 200\,au \citep[][scaled to GAIA DR2 distance]{2015ApJ...812..134H}.
33\,GHz continuum observations of FU~Ori with the Very Large Array (VLA) give a lower limit on the system dust mass of 8--16$\times10^{-5}\,\textnormal{M}_\sun$ 
with assumed brightness temperature 210$^{+81}_{-51}$\,K and opacity 0.07--0.13\,cm$^{2}$g$^{-1}$
\citep{2017A&A...602A..19L}.
The respective dust mass lower limits of FU~Ori and its companion FU~Ori~S, derived from ALMA continuum emission, are $2\times10^{-4}\textnormal{M}_\sun$ and $8\times10^{-5}\textnormal{M}_\sun$ \citep{2015ApJ...812..134H}.

\subsection{MWC~789 (HD~250550)}

MWC~789 is a Herbig Be star of spectral type B9
located in the Gemini OB1 molecular cloud complex
at a distance $697^{+94}_{-64}$\,pc and effective temperature of 8700\,K \citep{1995ApJ...450..201C,2004AJ....127.1682H,2016A&A...595A...1G,2018A&A...616A...1G,2018A&A...620A.128V}.
A stellar age of $1.42^{+2.21}_{-1.09}$\,Myr has been derived from VLT spectra, broadly consistent with the age of $2.56^{+0.30}_{-0.14}$\,Myr from fits to pre-main-sequence evolutionary tracks \citep[][\edit{respectively}]{2015MNRAS.453..976F,2018A&A...620A.128V}.

The object's H$\alpha$ polarization has a position angle of 175$^\circ$ 
\edit{which is unstable with a variation of 65$^\circ$},
and the continuum polarization is observed to vary on a month timescale \citep{1995A&AS..111..399J,2005MNRAS.359.1049V}.
This could be linked to the system's rotating accretion disk 
and fast stellar rotation shown by the short-term variability of the Ca\,\textsc{ii}\,K line \citep{1991A&A...244..166C,2002MNRAS.337..356V}.
The ratio of infrared excesses indicates a disk that is on the border between predicted transition disk and full disk structure (see Figure~\ref{fig:colourcolour}).
\edit{An SED fit, based on the model grid  of \cite{2007ApJS..169..328R}, gives a total disk mass $\log{(\textnormal{M}_{\textnormal{disk}})}=-1.02\pm0.13\log{(\textnormal{M}_\sun)}$
\citep{2011ApJ...734...22L}.}
There are no published scattered light images of this object, but \textit{H}-band NIR interferometry with VLTI/PIONIER has been used to derive a disk inclination of 52$^\circ$ and position angle $132^\circ$
\citep{2017A&A...599A..85L}. 
However we note that observations of CO ro-vibrational lines and the narrow OH $^{2}\Pi_{3/2}$ P4.5 (1+,1-) line using CRIRES on the VLT indicate a more face-on configuration with inclination 8--15$^\circ$ \citep{2011ApJ...732..106F,2016A&A...590A..98H}.

\edit{There have been 1.3mm continuum emission non-detections of both the circumstellar dust of the star and its associated globule CB39 with the IRAM 30\,m and SEST 15\,m telescopes, giving a gas mass \textless\,4.0$\times10^{-1}\,$M$_\sun$ \citep{1998A&A...336..565H,1997A&A...326..329L}. This is consistent with the \cite{2011ApJ...734...22L} disk mass estimate. \cite{2011ApJ...734...22L} also  report detections of emission lines of $^{13}$CO (2--1), CO (2--1), CO (3--2). The object also displays Brackett$-\gamma$ emission, indicative of an accretion rate of $1.6\pm0.4 \times 10^{-8}$\,M$_\sun$yr$^{-1}$ \citep{2011AJ....141...46D}. Ultraviolet absorption line observations have been used to estimate an H$_2$ column density of $1.81^{+0.89}_{-1.09}\times10^{19}\textnormal{cm}^{-2}$ \citep{2003A&A...410..175B} with radial velocity measurements suggesting it is bound to the star. We note that this H$_2$ column is consistent with the the line-of-sight to MWC~789 that probes a low-density envelope rather than a disk, suggesting the object has a low inclination.}


\subsection{HD~45677 (FS CMa)}

HD45677 is a variable Herbig Be star 
exhibiting \edit{the} B[e] phenomenon \citep{1976A&A....47..293A,1994A&AS..104..315T}.
%
%
The evolutionary status of HD~45677 is contested owing to its peculiar properties
such as forbidden line emission \citep{1976A&A....47..293A,1998A&A...340..117L}. 
These have lead to the creation of the FS~CMa classification of B[e] stars, of which HD~45677 is the prototypical object \citep{2007ApJ...667..497M}.
Its location on the HR diagram points to it being on/near the main sequence \citep{2007ApJ...667..497M} 
but its classification as a Herbig Be star suggests that it is pre-main-sequence \citep{1994A&AS..104..315T}.  
The SED is consistent with a full disk structure, which may imply that the object is at an early evolutionary stage.
%
An alternate interpretation of HD~45677 is that it is an evolved close binary, as the irregular emission spectra variations match those from hot winds from later evolutionary stages \citep{2015ApJ...809..129M}. HD~45677 is also isolated from star-forming regions and has been considered unlikely to be pre-main-sequence on this basis \citep{1994ASPC...62....3H}. 
The latest GAIA data release gives a distance to HD~45677 of 620\,pc \citep{2016A&A...595A...1G,2018A&A...616A...1G}
and \citet{2018A&A...620A.128V} use this with pre-main-sequence stellar tracks to derive an age of 0.6\,Myr.
This young age favours the interpretation that HD~45677 is a pre-main-sequence object. 
In Section~\ref{sec:res_hd45677} we perform our own analysis of the stellar photometry of this object in an attempt to better constrain the stellar properties.


The innermost region of the disk has been modelled using a ring with radius $<$10\,mas in order to fit \textit{H}-band IOTA interferometeric visibilities \citep{2006ApJ...647..444M}. The ring is inclined at 43$^\circ$ and has a position angle of 70$^\circ$ with the westward side of the ring facing towards the observer \citep{2017A&A...599A..85L}. 

Various studies have pointed towards HD~45677 having a bipolar outflow of material. 
An outflow: is consistent with complex Balmer lines \citep{1973A&A....26..443S, 1996A&A...311..643I}; is required for the continuous production of small grains responsible for the object's gradual color change \citep{1997A&AS..121..275D}; and is necessary to explain the variation with wavelength of the degree and position angle of polarization \citep{1992ApJ...401L.105S}.  \citet{2006MNRAS.373.1641P} propose that HD~45677 has a polarization angle of $164\pm3^\circ$ and a outflow with a position angle of $175\pm1^\circ$.
%


\subsection{Hen~3-365 (HD~87643)}

Hen~3-365 is a system containing a massive star that exhibits B[e] phenomenon and is thought to have a mass of $\sim$25\,M$\sun$ \citep[e.g.][]{1998MNRAS.300..170O}.
\textit{K} and \textit{H}-band image reconstructions from VLTI/AMBER reveal a companion with a separation of $\sim$34$\pm$0.5mas \citep{2009A&A...507..317M}.
The two stars are surrounded by an oxygen-rich dusty envelope or circumprimary dust ring at ~6\,au, with an inner dusty disk radius $\sim$2.5-3.0au \citep{2009A&A...507..317M}.

The distance to Hen~3-365 is uncertain and contributes to the unknown evolutionary status of the star. 
The extinction and kinematics indicate that Hen~3-365 is more likely an evolved supergiant at a distance of several kpc than a close young star
\citep{1998MNRAS.300..170O},
and \citet{2018MNRAS.480..320M} classify Hen~3-365 as a massive supergiant.
However the HI column density is far too high for Hen~3-365 to be a supergiant \citep{1998MNRAS.300..170O}. 
\citet{2018A&A...620A.128V} derive a distance to Hen~3-365 of 2010\,pc using a prior, as
the GAIA DR2 distance to Hen~3-365 is unreliable due to a poor parallax calculation (potentially caused by confusion with the object's close companion).
But \citet{2018A&A...620A.128V} note that if Hen~3-365 is an evolved object, 
it would be inappropriate to use pre-main-sequence tracks as they have.
This could partially explain the large uncertainties on the other derived parameters for Hen~3-365 in Table~\ref{tab:masses}.

Assuming that Hen~3-365 is a pre-main-sequence object, the SED of Hen~3-365 is best modelled using a transition disk structure 
as the mid-infrared excess is greater than the NIR excess (see \edit{Table~\ref{tab:mags}}, Figure~\ref{fig:colourcolour}).
The system displays 1.0--2.5$\mu$m NIR excess from thermal dust emission
with a hot and cold dust mass of $1.6\times10^{-7}$\,M$_\sun$ and $3.7\times10^{-3}$\,M$_\sun$ respectively \citep{1973MNRAS.161..145A,1988ApJ...324.1071M}.
CO band emission is observed in a ring $\sim$200\,au from the star,
and is best fit with a model ring with a 
near pole-on inclination of 7.4$^{\circ}$ \citep{2018MNRAS.480..320M}.

The Hen~3-365 system contains a fast polar wind, \edit{a} slow disk wind, and \edit{a} large nebula that impact the disk properties \citep{1998MNRAS.300..170O}.
The reflection nebula is seen in cool dust emission 
and has broken structures that could be due to periodic mass transfer between the two stars if Hen~3-365 is an evolved system
\citep{1988ApJ...324.1071M,2009A&A...507..317M}.
The geometry of these nebula components matches the smaller-scale polarization angle of the asymmetric disk seen with spectropolarimetry \citep{1998MNRAS.300..170O}.
%
The reflection nebula is likely caused by mass loss via an outflow as seen from P Cygni profiles of Ca\,\textsc{ii} H and K and Balmer lines \citep{1981A&A....93..285S}.
The outflow velocity is variously described as having a lower limit of 1200\,km/s 
or a velocity of $\sim$\,800km/s \citep{1981A&A....93..285S,1985A&A...152..101D}.
\citet{2006MNRAS.367..737B} describe two outflows, one symmetric in the EW direction and one asymmetric in the N direction, that could have been caused by a bow shock structure in the north.
%

\subsection{HD~139614}

HD~139614 is an A7-type star in the Sco OB2-3 association whose age estimates range between 9\,Myr and 14.5\,Myr \citep{2013MNRAS.429.1001A,2014A&A...561A..26M,2004A&A...426..151A,2018A&A...620A.128V}
with a stellar mass of 1.8\,M$_{\sun}$ 
and accretion rate 10$^{-8}$\,M$_{\sun}$~yr$^{-1}$ \citep{2006A&A...459..837G,2013MNRAS.429.1001A}.
The BANYAN $\Sigma$ association finder places HD~139614 in the Upper Centaurus-Lupus group with 99.9\% likeliness \citep{2018ApJ...856...23G}. This is a group with stars of ages 12--15Myr \citep{1989A&A...216...44D}, and \citet{2018A&A...620A.128V} derive an age of HD~139614 of 14.5\,Myr from pre-main-sequence tracks. 
The object has a transition disk structure based on its infrared excesses (see \edit{Table~\ref{tab:mags}}, Figure~\ref{fig:colourcolour}).
%
The warm dust mass is $2.9\times10^{-8}$\,M$_\sun$ \citep{1998A&A...329..131M,2011ApJ...737...57M}, and
the disk contains polycyclic aromatic hydrocarbons with a mass within 10\,au of the star of $1.83\times10^{-6}$\,M$_{\rm Earth}$ \citep{2017ApJ...835..291S}.
%
The NIR-emitting region of the disk spans 0.2--2.5au with a dust scale-height of $\sim$0.01\,au at a radius of 0.2\,au \citep{2016A&A...586A..11M}.
There is a gap between \edit{an inner halo component} and dust disk 
from 2.5--5.7\,au with a $10^3$ factor of depletion compared with the outer disk \citep{2014A&A...561A..26M,2015A&A...581A.107M,2016A&A...586A..11M},
which continues from 5.6\,au to an outer radius of 75--100\,au \citep{2009A&A...508..707P,2014SPIE.9146E..2TL}.

The object's very low $v\sin(i)$ (24$\pm1\textnormal{kms}^{-1}$),  \citep{2013MNRAS.429.1001A,1997MNRAS.290..165D} is indicative of a near pole-on inclination of $\sim5^\circ$. This is supported by the object's low linear polarization \citep{1999A&A...345..547Y}.
disk model fits to NIR VLTI visibilities reveal that the inner disk is discontinuous although there are no significant brightness asymmetries \citep{2014A&A...561A..26M,2016A&A...586A..11M}, 
and the inner rim is optically thin and not puffed up, allowing the star to warm the inner wall of the outer disk \citep{2001A&A...365..476M,2014A&A...561A..26M}.
Warm dust in the flared outer disk is thought to cause the disk's mid-IR emission in the SED \citep{2001A&A...365..476M}.

Hydrodynamical simulations suggest that a planetary companion of mass  $\sim$3M$_{\rm Jup}$ may be able to open the gap at 4.5au from the star \citep{2016A&A...586A..11M}.
A planetary companion at 4au with mass $<2 \rm M_{Jup}$ is also consistent with model fits to CO lines \citep{2017A&A...598A.118C}. 

There has been an \textit{H}-band scattered light non-detection by Subaru-HiCIAO using adaptive optics and a coronagraph but without polarimetry \citep{2010PASJ...62..347F}. 
Other Subaru-HiCIAO \textit{H}-band polarimetric differential imaging observations found no companion candidates within 400\,au \citep{2017AJ....153..106U}. 




\section{Observations and data reduction}
\label{sec:observations}
\subsection{Observations}

\begin{table*}
	\centering
	\caption{
	Details of our observations.
    }
	\label{tab:observations}
	\begin{tabular}{lllllllll}
		\hline
		Object & Midpoint (UTC) &  Band & $t_{int}$ (s) & Coadds & Frames & UTOS & Airmass & \edit{RMS (nm)}\\
		\hline 
        FU~Ori    & 2018-01-03 05:16:00 & \textit{J} & 29.10 & 2 & 24 & 23\,m\,16.8\,s & 1.4--1.5 & $216 \pm 9$ \\
        MWC~789   & 2018-11-20 08:50:05 & \textit{H} & 29.10 & 2 & 32 & 31\,m\,\phantom{0}2.4\,s & 1.6--1.9 & $271 \pm 6$ \\
        HD~45677  & 2017-12-31 05:16:45 & \textit{J} & 29.10 & 2 & 32 & 31\,m\,\phantom{0}2.4\,s & 1.1 & $161 \pm 6$ \\
        HD~45677  & 2018-01-01 06:54:50 & \textit{H} & 29.10 & 2 & 32 & 31\,m\,\phantom{0}2.4\,s & 1.2 & $199 \pm 11$ \\
        Hen~3-365 & 2017-04-06 03:27:39 & \textit{J} & 14.55 & 4 & 28 & 27\,m\,\phantom{0}9.6\,s & 1.2 & $147 \pm 9$ \\
		HD~139614\tablenotemark{a} & 2017-04-06 07:57:26 & \textit{J} & \phantom{0}2.91 & 10 & 16 & \phantom{0}7\,m\,45.6\,s &1.0 & $165 \pm 8$ \\
		HD~139614 & 2017-04-06 09:06:56 & \textit{J} & 29.10 & 2 & 28 & 27\,m\,\phantom{0}9.6\,s & 1.1 & $182 \pm 14$ \\
        HD~139614 & 2018-06-08 00:40:17 & \textit{H} & 29.10 & 2 & 24 & 23\,m\,16.8\,s & 1.2 & $256 \pm 10$ \\
		HD~139614 & 2019-05-13 05:31:45 & \textit{J} & 29.10 & 2 & 32 & 31\,m\,\phantom{0}2.4\,s & 1.0 & $196 \pm 5$ \\
		\hline
	\end{tabular}
	{\par Notes:
	We present UTC date and time at midpoint of exposure,
    the filter \textit{J} or \textit{H},
    $t_{int}$ integration time or exposure time per frame in seconds,
    number of co-added images,
    number of frames combined into the final images,
    total Usable Time On Source (UTOS) accounting only for used frames,
    airmass range throughout the observation,
    and \edit{mean RMS wavefront error with standard deviation} across the whole observation.
	}
	\tablenotetext{a}{Without coronagraph.}
\end{table*}

Our observations were taken using the Gemini Planet Imager (GPI) \citep{2014PNAS..11112661M}, an instrument on the Gemini South telescope, during 2017, 2018, and 2019. 
The details including observation dates and data quality are in Table~\ref{tab:observations}. 
The program IDs containing our observations are: GS-2017A-LP-12, GS-2017B-LP-12, GS-2018A-LP-12, GS-2018B-LP-12, and GS-2019A-LP-12.

All observations used GPI in polarimetry mode and wavebands \textit{J} or \textit{H}. 
For most targets individual exposures had an integration time of 29.10s, with the exception of Hen~3-365 and HD~139614 (without spot) with shorter times of 14.55s and 2.19s respectively to prevent or minimise saturation.
Our objects typically had 32 usable exposures resulting in approximately one hour total observing time per target including overheads. 
Files were determined to be unusable if they suffered from poor adaptive optics performance.  
All of our data were taken in 70th-percentile image quality, the second-best possible tier and translating to seeing of 0.4--0.7" for \textit{J}-band and 0.4--0.5" for \textit{H}-band.

Our observations use Polarimetric Differential Imaging (PDI); \citep[e.g.][]{2015ApJ...799..182P}.
A half-wave plate (HWP) is used to restrict the polarization position angle of light that passes through the instrument.
We use four HWP angles (0$^\circ$, 22.5$^\circ$, 45$^\circ$, 67.5$^\circ$);
note: the April 2017 observations use an additional four angles
where 90$^\circ$, 112.5$^\circ$, 135$^\circ$, 157.5$^\circ$ are treated as equivalent to 0$^\circ$, 22.5$^\circ$, 45$^\circ$, 67.5$^\circ$ respectively.
After the HWP the light passes through a polarizing Wollaston prism
and is split into two beams with orthogonal polarization angles
that fall onto the detector simultaneously.
Frames from four HWP angles can be combined into a \edit{Stokes data cube (stokesdc - see following Section~\ref{sec:reduction}}
using singular value decomposition and double-differencing \citep[e.g.][]{2015ApJ...799..182P}.

The observations use an Apodized-Pupil Lyot Coronagraph with spot radius 0.085" for \textit{J}-band and 0.111" for \textit{H}-band images \citep{2014PNAS..11112661M}.
\edit{A coronagraph was used in all cases except for observations of HD~139614 in \textit{J}-band from 2017, where a set of observations was taken both with and without a coronagraph.}
The pixel scale of the detector is 0.0141'' giving each stokesdc a maximum field of view of 2.7''$\times$2.7''.

\subsection{Data reduction}
\label{sec:reduction}

\begin{table*}
	\centering
	\caption{
    	Calculated scale factors for converting data cube units from counts (ADU/coadd)/s to surface brightness (mJy/arcsec$^2$).
    }
	\label{tab:fluxcalib}
	\begin{tabular}{llllll}
        \hline
        Object	&	Band	&	Magnitude	&	Calibfac 	&	Caliberr	&	Scale factor    \\
                &           &               &   (10$^{-8}$) &     (\%)      & (mJy/arcsec$^2$)/ \\
                &           &               &               &               & [(ADU/coadd)/s]  \\
        \hline
        FU~Ori\tablenotemark{a}	    &	\textit{J}	&	6.519	$\pm$	0.023	&  4.5 & 11 & $2.2\pm0.5$ \\
        MWC~789     &   \textit{H}   &   7.528   $\pm$   0.026   &   2.6 &  $\phantom{0}$6.7 &  3.6 $\pm$ 0.5    \\
        HD~45677	&	\textit{J}	&	7.242	$\pm$	0.026	&	1.6 &  $\phantom{0}$8.3 &  1.7 $\pm$ 0.4    \\
        HD~45677	&	\textit{H}	&	6.347	$\pm$	0.023	&	2.5 &  $\phantom{0}$9.3 &  3.5 $\pm$ 0.6    \\
        Hen~3-365	&	\textit{J}	&	6.217	$\pm$	0.037	&	5.0 &  $\phantom{0}$8.0 &  2.6 $\pm$ 0.6    \\
        HD~139614	&	\textit{H}	&	7.333	$\pm$	0.040	&	2.2 &  $\phantom{0}$8.8 &  3.1 $\pm$ 0.5    \\
        HD~139614	&	\textit{J}	&	7.669	$\pm$	0.026	&	1.7 &  $\phantom{0}$6.2 &  1.8 $\pm$ 0.4    \\
        HD~139614 (2017)	&	\textit{J}	&	7.669	$\pm$	0.026	&  2.7 & $\phantom{0}$8.4 &  2.8 $\pm$ 0.6    \\
        \hline										
	\end{tabular}
	{\par Notes:
    All magnitudes are from the 2MASS survey \citep{2003yCat.2246....0C}. 
    ``Calibfac" and ``caliberr" are the two parameters returned by flux calibration primitives in the GPI data reduction pipeline.
    We convert these parameters into the final ``Scale factors" presented.
	}
	\tablenotetext{a}{FU~Ori has an inconsistent scale factor ($4.8\pm1.1$) and so instead we use the mean value of the other \textit{J}-band objects' scale factors, $2.2\pm0.5$ (see Section~\ref{sec:reduction}).}
\end{table*}

Our data reduction follows the process in \citet{2017ApJ...838...20M,2019ApJ...872..122M} using the GPI Data Reduction Pipeline (DRP) version 1.4 routines to convert between file types - raw data files, polarization data cubes (podc), and Stokes data cubes (stokesdc) \citep{2010SPIE.7735E..31M}. 
We also perform stellar polarization corrections and combine multiple data cubes using original routines written in python to manipulate output files from the GPI DRP.
The data were calibrated using dark and flat frames taken near the respective observation dates, which in turn were calibrated using the DRP as per the documentation. 
We note that we did not calibrate using low spatial frequency flat fields.
\edit{When reducing the data, we did not include the system Mueller matrix when converting from podc to Stokes in the combine Polarization Sequence primitive.}

We follow conventions for the Stokes parameters $Q$, $U$ and the radial Stokes parameters $Q_\phi$, $U_\phi$ given in \citet[][Appendix A]{2019ApJ...872..122M}. 
In short, $Q$ and $U$ follow the IAU standards \citep{1974IAU} and $Q_\phi$, $U_\phi$ use a modified version of that defined in \citet{2006A&A...452..657S}, where now azimuthally-oriented polarization is positive and polarization in the radial direction is negative.
As such, the azimuthal Stokes maps $Q_\phi$ and $U_\phi$ have the following form:
\begin{equation}
Q_{\phi} = - U\sin(2\phi) - Q\cos(2\phi)
\label{eqn:qphi}
\end{equation}
\begin{equation}
U_{\phi} = +Q\sin(2\phi) - U\cos(2\phi)
\label{eqn:uphi}
\end{equation}
with polar angle 
\begin{equation}
\phi = \tan^{-1} \left(\frac{Y-Y_{0}}{X-X_{0}} \right) + \gamma
\end{equation}
for the pixel grid of X and Y coordinates with center pixel (X$_0$,Y$_0$) and the position angle $\gamma$. 

The basic steps of the reduction process are as follows:
\begin{enumerate}
\item DRP: raw data converted to podc
\item DRP: each group of four podc converted to stokesdc
\item Python routines: stellar polarization correction for each stokesdc
\item Python routines: all stokesdc combined into one final stokesdc
\end{enumerate}

To begin the reduction, the raw data files were converted into podc containing two orthogonally polarized images using the standard DRP recipe ``Simple Polarization data cube Extraction".
A dark field was subtracted from each image, 
then a flat field was used to find the expected locations of light on the detector. 
Finally the images were destriped, had bad pixels interpolated, and were assembled into a podc.

The podc were split into groups of four podc that contained a full HWP cycle.
The podc were checked individually by eye for evidence of poor adaptive optic (AO) performance and cycles containing poor files were removed.
Podc with poor AO performance had unusually extended point spread functions (PSFs) \edit{and very weak AO spots.}
\edit{These frames typically have larger RMS wavefront error values compared to frames with better AO of the same object. For example, FU~Ori frames that were removed for poor performance had larger RMS wavefront values of $\sim$\,260 compared to to frames with better AO performance of  $\sim$\,216.}
When a poor podc was identified, the whole HWP cycle containing the podc was removed.
We removed the final two cycles for FU~Ori, the first cycle for Hen~3-365, and the final two cycles of HD~139614 from 2017 with the coronagraph. These changes are reflected in Table~\ref{tab:observations} \edit{along with the average RMS wavefront error values for that observing epoch.}.

Each group was reduced separately into a stokesdc using the standard DRP recipe ``Basic Polarization Sequence", without the step to subtract mean stellar polarization.
In this recipe the polarization pairs are cleaned using double differencing. Then all images were rotated to North and aligned,
and converted to a stokesdc. We use the term ``interim stokesdc'' to refer to these stokesdc that are formed from only one HWP cycle. Later in the reduction process the multiple interim files will be combined into one final stokesdc.

We then apply the stellar polarization correction to each of the multiple interim stokesdc.
We found that the correction method used could give noticeably different resulting radial Stokes maps, particularly in the lower-quality $U_\phi$ maps.
\edit{
Similarly to \citet{2019ApJ...872..122M} we expect effects from instrumental polarization to be non-dominant (with systematics on the order of 1\% of total intensity and a few percent for $Q_\phi$) and so we have not taken further steps to remove these effects outside the following. 
}
We compared results from the DRP with other methods including $U_\phi$ minimisation \citep[e.g.][]{2018ApJ...863...44A} and fractional total intensity subtractions in Appendix~\ref{app:reduction_methods}.
Most of the methods could not completely remove the contribution from stellar polarization, which causes a ``butterfly'' pattern in $Q_\phi$ and $U_\phi$ that can obscure structures and change their sign.
For the objects presented in this paper, we have elected to use the method that consistently removes the background stellar polarization while introducing few structures in $Q_\phi$ and $U_\phi$ compared with the DRP reduction.
The method works as follows:
in each interim Stokes cube, we examine a ring-shaped region near the detector's edge with radius $70\leq r\leq80$ pixels ($\sim1.0\leq r\leq1.1$'').
Using all pixels centered in this region we calculate ratios of Stokes $Q$ and $U$ to total intensity $I$, $f_Q$ and $f_U$ respectively:
\begin{equation}
    f_Q = \frac{\textnormal{mean(}Q\textnormal{)}}{\textnormal{mean(}I\textnormal{)}}
\end{equation}
\begin{equation}
    f_U = \frac{\textnormal{mean(}U\textnormal{)}}{\textnormal{mean(}I\textnormal{)}}
\end{equation}
These ratios are used as scaling factors across the full map to calculate weighted Stokes maps $Q^\star$ and $U^\star$:
\begin{equation}
    Q^\star = Q - I\times f_Q
\end{equation} 
\begin{equation}
    U^\star = U - I\times f_U
\end{equation}
Then the corrected stokesdc is saved, containing the images $[I,Q^\star,U^\star,V]$.
This gives radial Stokes maps with this form: 
\begin{equation}
Q_\phi = -U^{\star}\sin(2\phi) - Q^{\star}\cos(2\phi)
\end{equation}
\begin{equation}
U_\phi = +Q^{\star}\sin(2\phi) - U^{\star}\cos(2\phi)
\end{equation}

Finally all corrected interim stokesdc were combined into one final stokesdc. This combination used the mean of all maps for each Stokes parameter, e.g. $Q_{\rm final} = <Q_1,Q_2,...,Q_n>$ for $n$ interim stokesdc.
Conversely the default DRP method creates one stokesdc from all $n\times4$ input files and performs the stellar background removal once on the resulting stokesdc.
Combining the interim stokesdc in this way 
results in an improved signal-to-noise ratio and works to counter the effects of the rotation of the sky relative to the fixed detector, which can cause any unobscured stars on the frame (e.g. companions) to appear smeared-out over an hour-long set of exposures. 
Reducing the files in smaller batches essentially limits the exposure to a matter of minutes and allows less time for the companion star to move relative to the detector. 
This effect is particularly pronounced for companion stars owing to their much higher flux relative to background levels on the frame.
Since the data is averaged without correcting for sky rotation, the PSFs add together in the detector frame of reference.

Flux calibration was performed using standard GPI DRP routines to measure the satellite spot flux, as in \citet{2019ApJ...872..122M}. For each object all podc were stacked using the primitives ``Accumulate Images" and ``Combine 3D data cubes" to create a podc made from the mean of all other cubes without rotating to north
to prevent effects from the rotation of the detector with respect to the sky, e.g. the satellite spots would no longer be in a predefined area and may have spread out depending on the duration of all frames,
\edit{as can be seen in Figure~\ref{fig:toti} where the spots of HD~139614 in \textit{J}-band appear stretched compared with those in FU~Ori.}
The satellite spot fluxes were then measured in this mean cube only
as it has higher signal-to-noise than the separate podc.
For \textit{J}-band images we used second-order satellite spots that are more free of Speckles.
These second-order satellite spots contain 25\% less flux than the first-order spots.
This factor is unaccounted for by the pipeline,
resulting in the returned calibration factors being too large.
To counteract this we apply an additional factor of 0.75 to the final flux scales for \textit{J}-band objects,
a change which is reflected in Table~\ref{tab:fluxcalib}.

The scaling factor is calculated from the known ratio of the stellar magnitude from Table~\ref{tab:background} and the satellite spot flux using the following primitives: ``Measure Star Position for Polarimetry"; ``Measure Satellite Spot Flux in Polarimetry"; ``Calibrate Photometric Flux in Pol Mode".
The scaling factor $X$ follows that used by 
\citet{2016ApJ...818L..15W}, where the DRP default units (Analog-to-Digital Units (ADU)/coadd) are equal to $X$ mJy s$^{-1} \cdot \textnormal{arcsec}^{-2}$. 
The scale factor $X$ is found by multiplying the initial scaling factor returned by the DRP by each frame's exposure time in seconds and dividing by the area of one pixel on the GPI detector, 0.0141arcsec$^{2}$.
The uncertainty in the flux scale values is dominated by the systemic error, 20\% for \textit{J}-band and 13\% for \textit{H} \citep[see][]{2019ApJ...872..122M}, once these values are combined in quadrature with the percentage error returned by the DRP.

Table~\ref{tab:fluxcalib} gives the scale factors for our observations.
The scale factor for FU~Ori is inconsistent for this sample as it is double the mean of the other \textit{J}-band values. This is likely due to an inaccurate or outdated 2MASS magnitude, which does not reflect the gradual fall in the intensity of FU~Ori since its initial flare-up in 1936 \citep{1966VA......8..109H}. Therefore we have taken the scale factor for FU~Ori to be this mean value, $2.2\pm0.5$.

\section{Results and Discussion}
\label{sec:results}

\begin{figure*}
	\centering
	\includegraphics[width=0.8\textwidth]{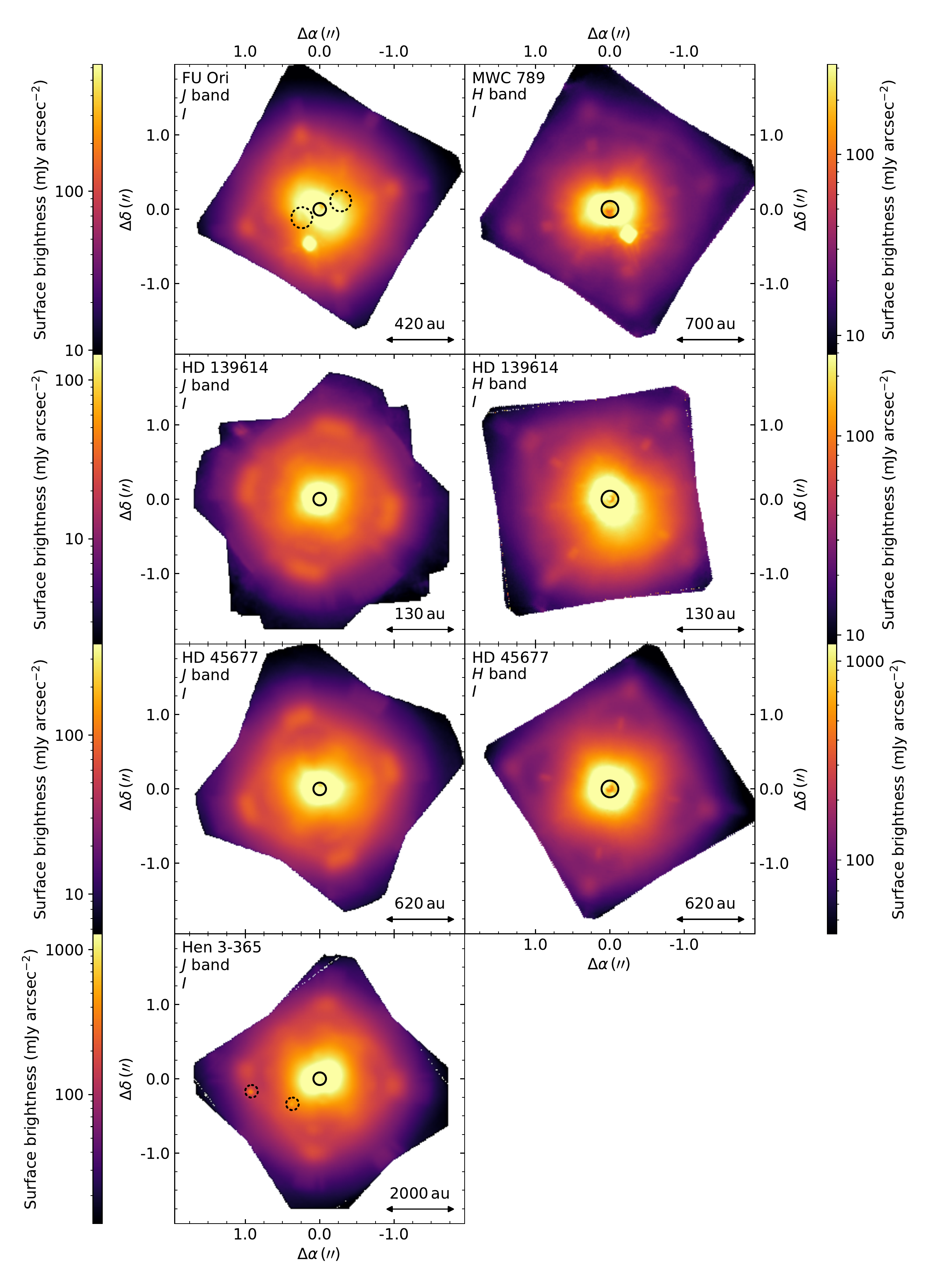}
    \caption{
    	Total intensity images for our observed targets. 
    	North is up, East is left.
    	The colorbars show surface brightness on arbitrary logarithmic scales.
    	Each double-ended arrow indicates the physical scale derived from the GAIA distances in Table~\ref{tab:masses}.
        The brightest region near the coronagraphic spot is the stellar PSF and can appear elongated due to strong wind shaking the telescope during the observations.
    	The bright spots in the four corners of each image are satellite spots. These appear at radii of 1.0'' in \textit{H}-band, and secondary satellite spots appear at 1.0'' in J-band and 1.5'' in \textit{H}-band. 
    	Companion candidates are visible around FU~Ori, MWC~789, and Hen~3-365, with their locations given in Table~\ref{tab:companions}.
        Other faint visible features include the arm to the north-west of MWC~789, the arm to the immediate east of the spot in FU~Ori, and the arm to the north west of the spot in Hen~3-365.
        \edit{The dashed circle annotations highlight companion candidates in Hen~3-365 and two faint ''spots" of unpolarised light in FU~Ori caused by a minor instrumental error during the observation (see Section~\ref{sec:results}).}
    }
    \label{fig:toti}
\end{figure*}

\begin{figure*}
	\centering
    \includegraphics[width=0.8\textwidth]{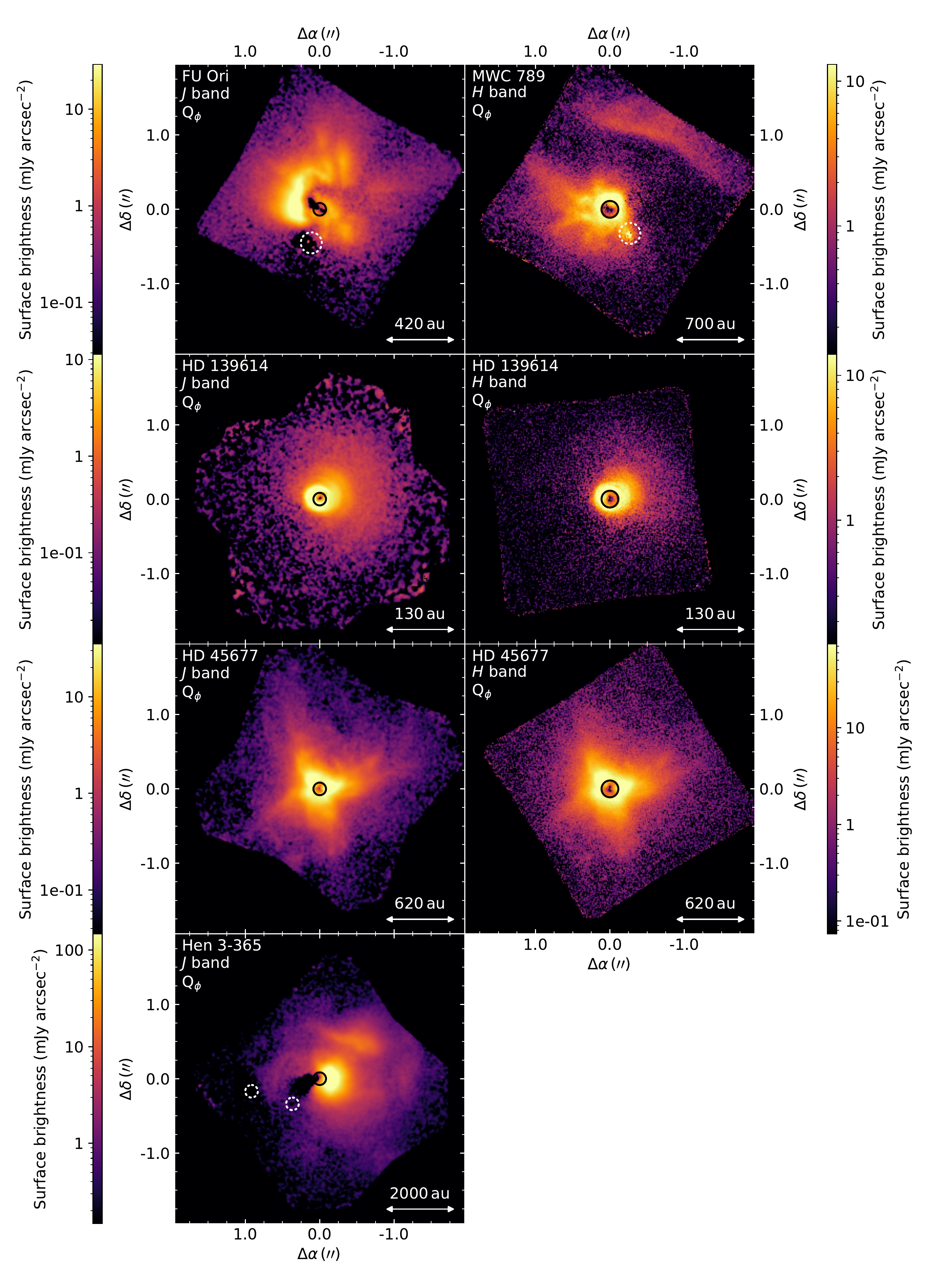}
    \caption{
    	Radial Stokes $Q_\phi$ images for our observed targets. 
    	North is up, East is left.
    	The physical scales are the same as in Figure~\ref{fig:toti}.
    	\edit{The dashed circle annotations mark all four companion candidates.}
    }
    \label{fig:qphi}
\end{figure*}

\begin{figure*}
	\centering
    \includegraphics[width=0.8\textwidth]{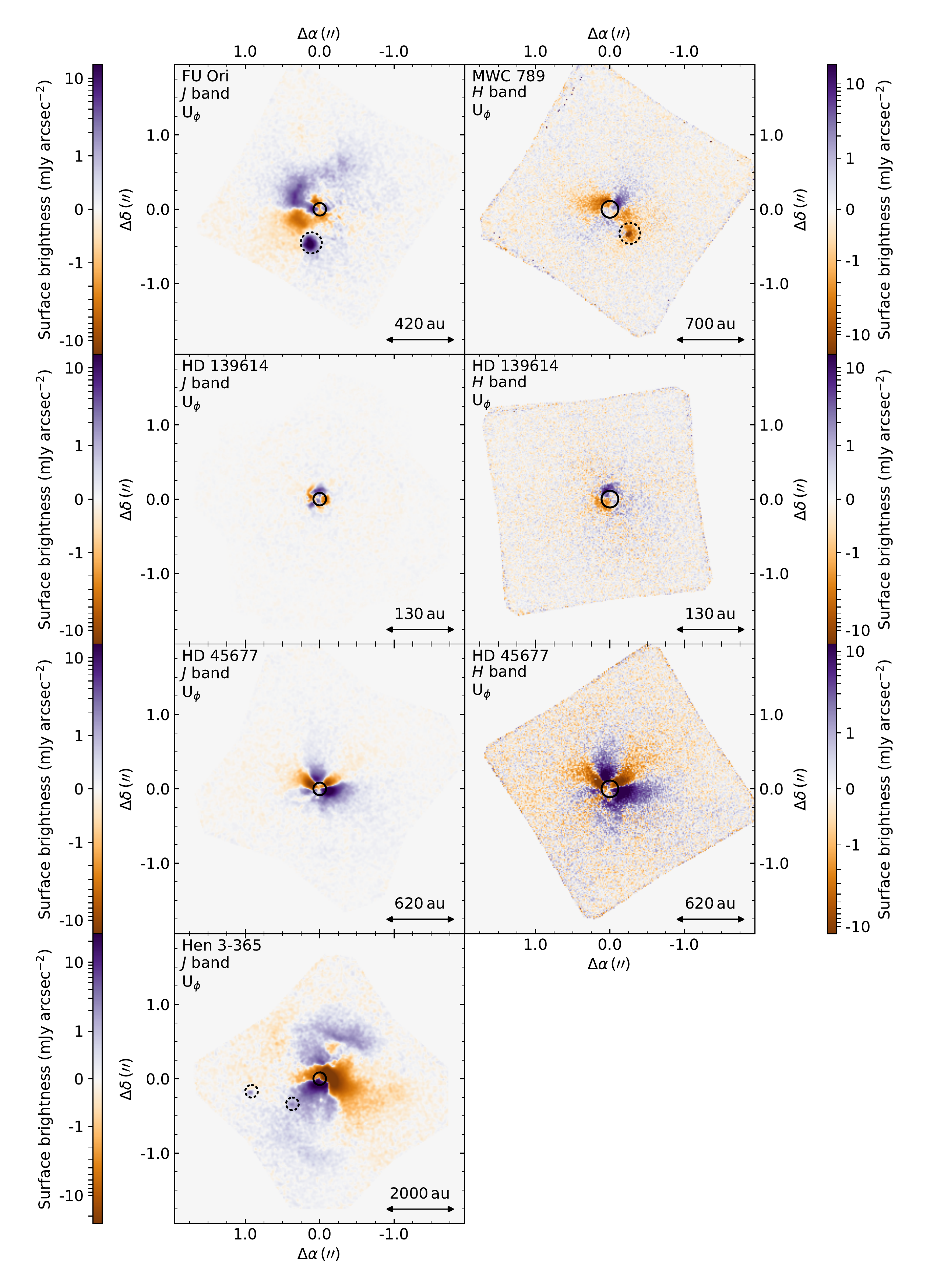}
    \caption{
    	Radial Stokes $U_\phi$ images for our observed targets. 
    	North is up, East is left.
    	The physical scales are the same as in Figure~\ref{fig:toti}.
    	\edit{The dashed circle annotations mark all four companion candidates.}
    }
    \label{fig:uphi}
\end{figure*}

\begin{figure}
	%
	\includegraphics[width=0.23\textwidth, trim=0cm 0 0cm 0cm, clip]{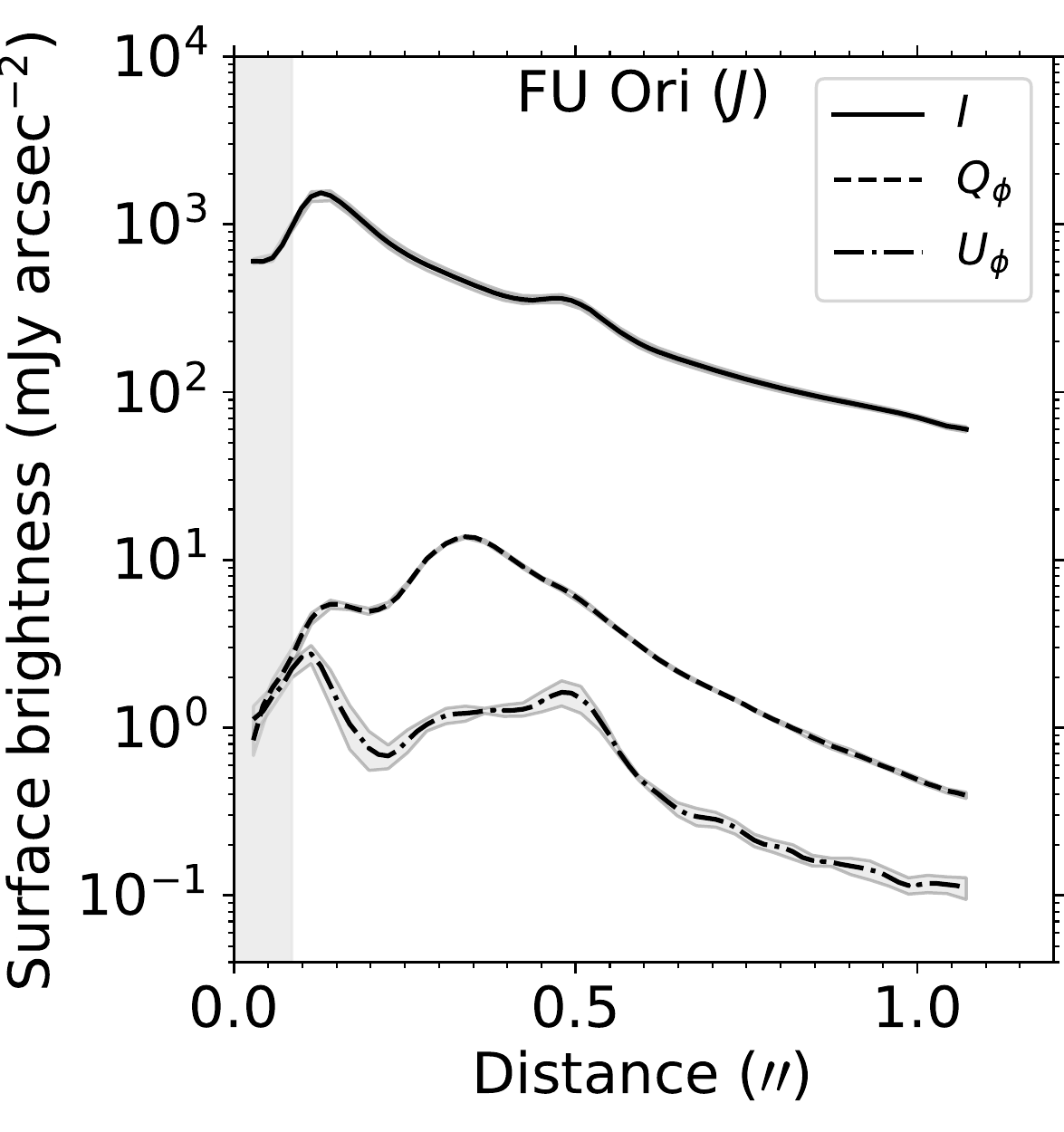}
	\includegraphics[width=0.23\textwidth, trim=0cm 0 0cm 0cm, clip]{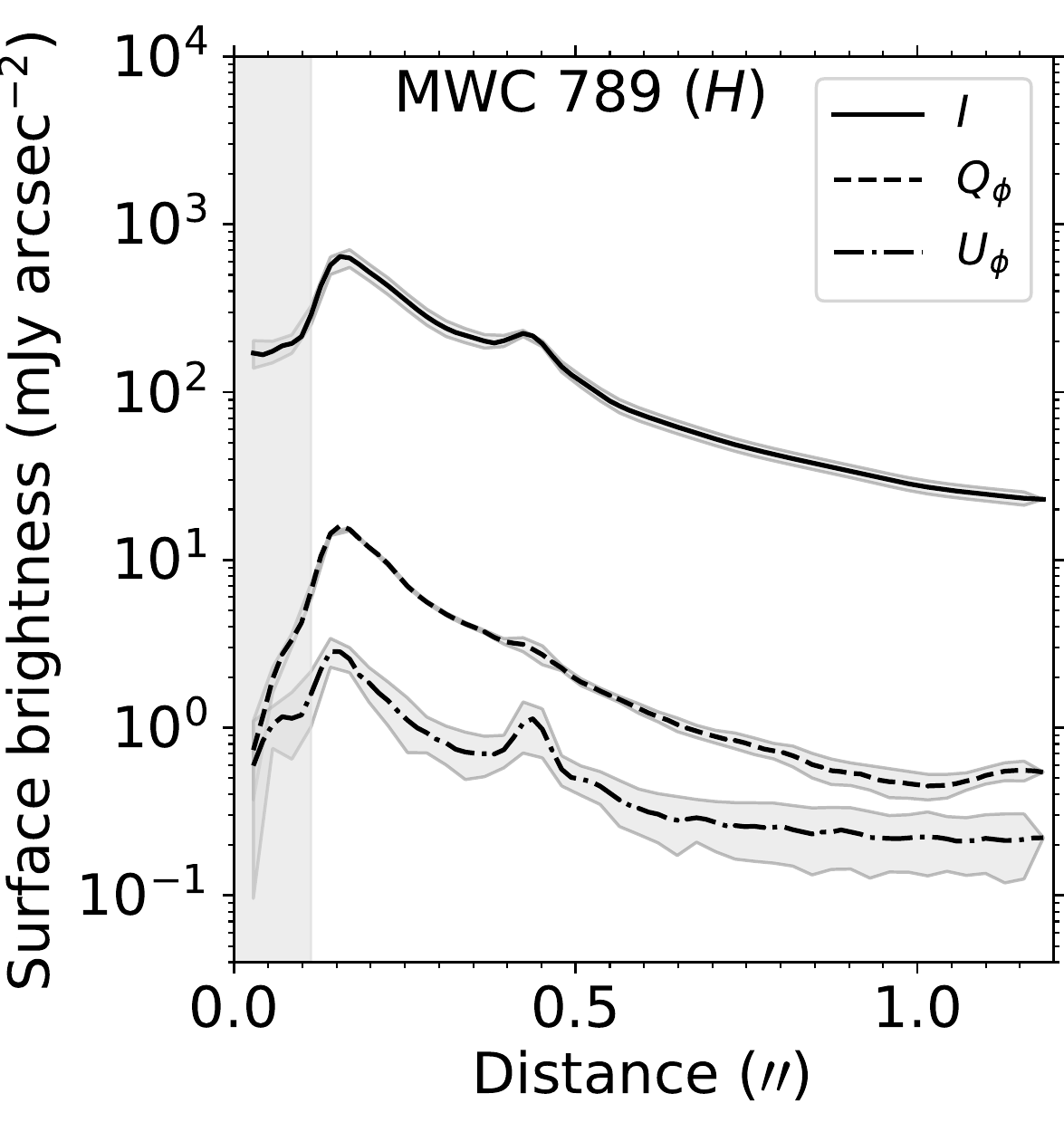}
	\includegraphics[width=0.23\textwidth, trim=0cm 0 0cm 0cm, clip]{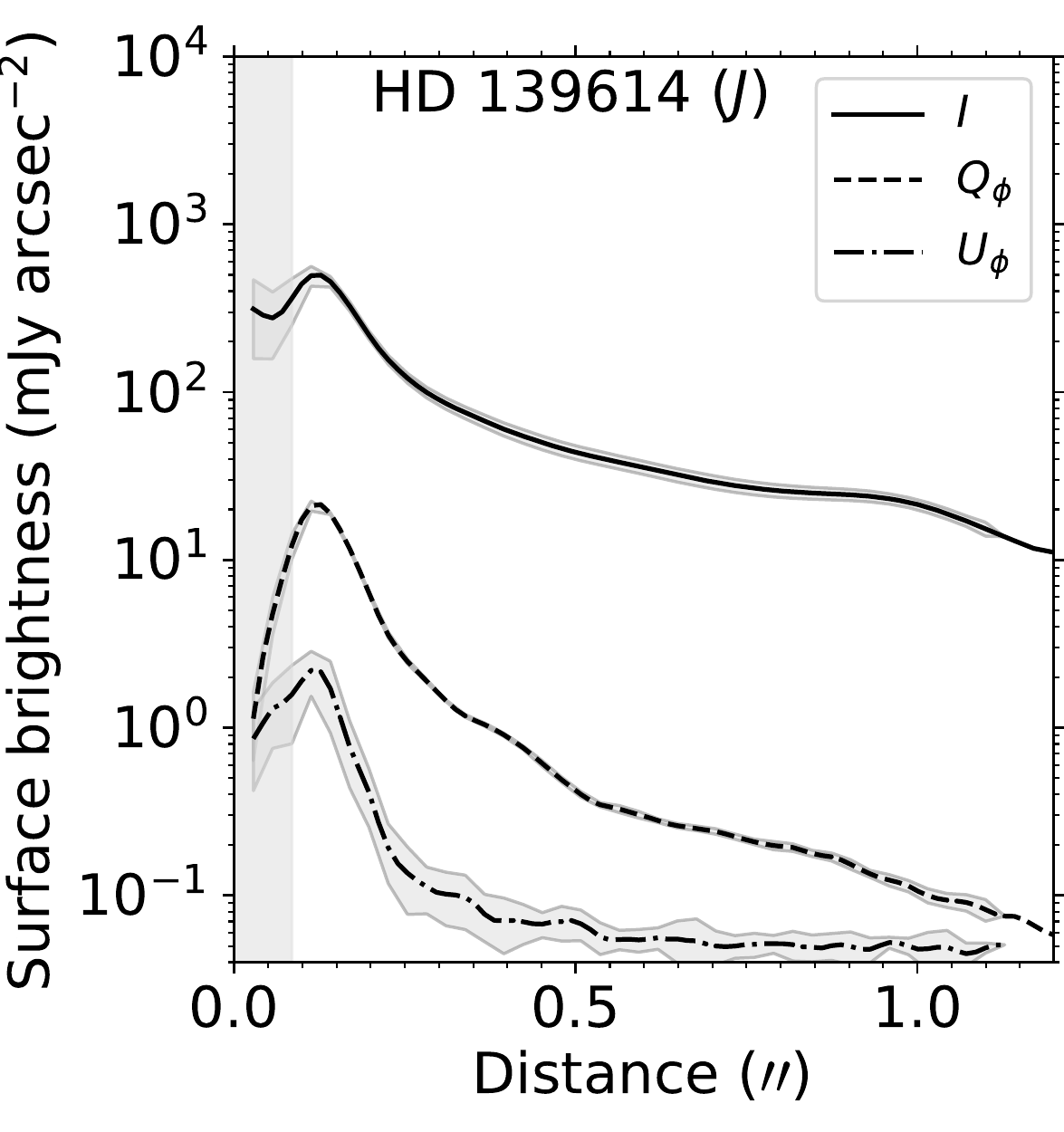}
	\includegraphics[width=0.23\textwidth, trim=0cm 0 0cm 0cm, clip]{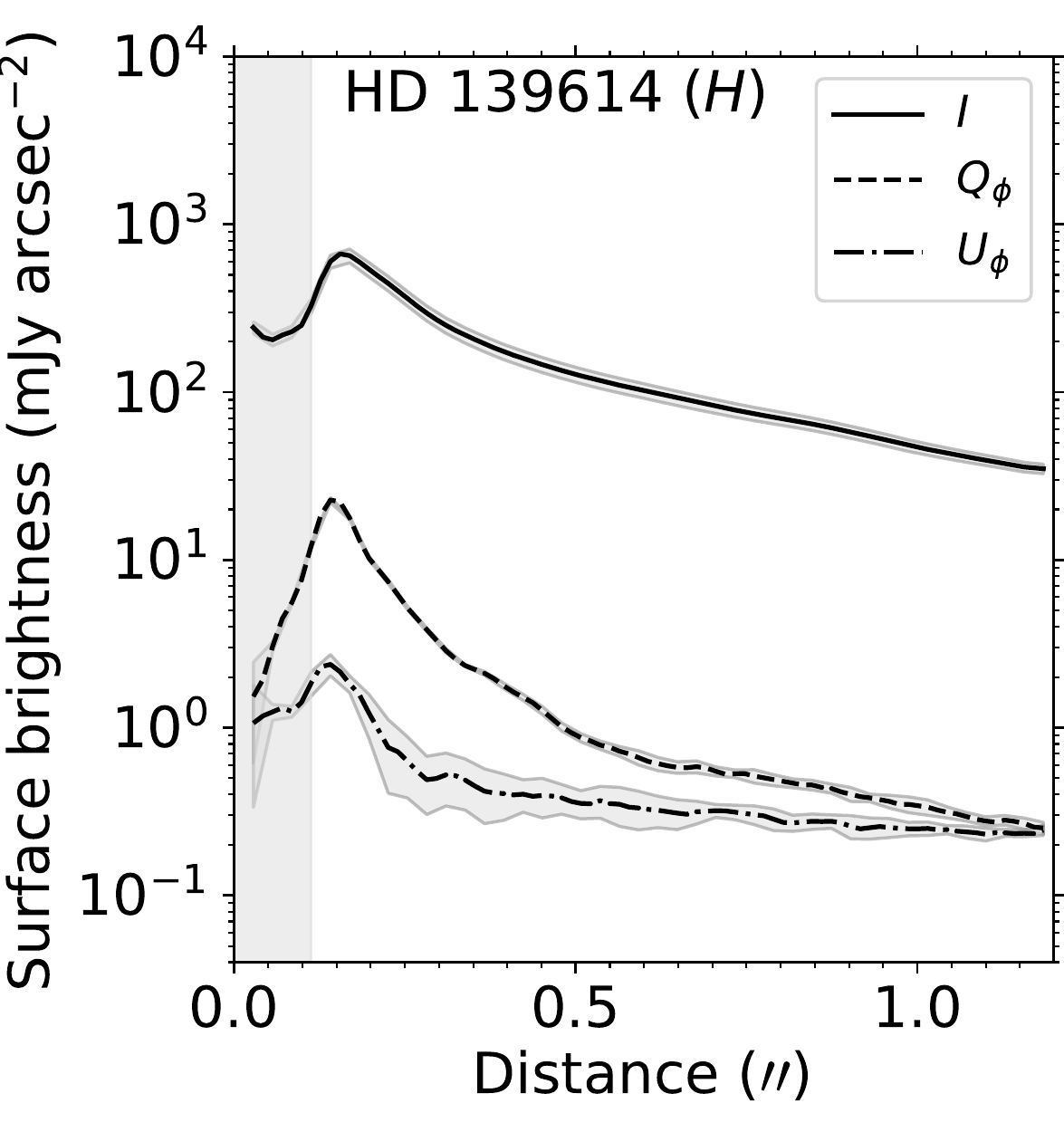}
	\includegraphics[width=0.23\textwidth, trim=0cm 0 0cm 0cm, clip]{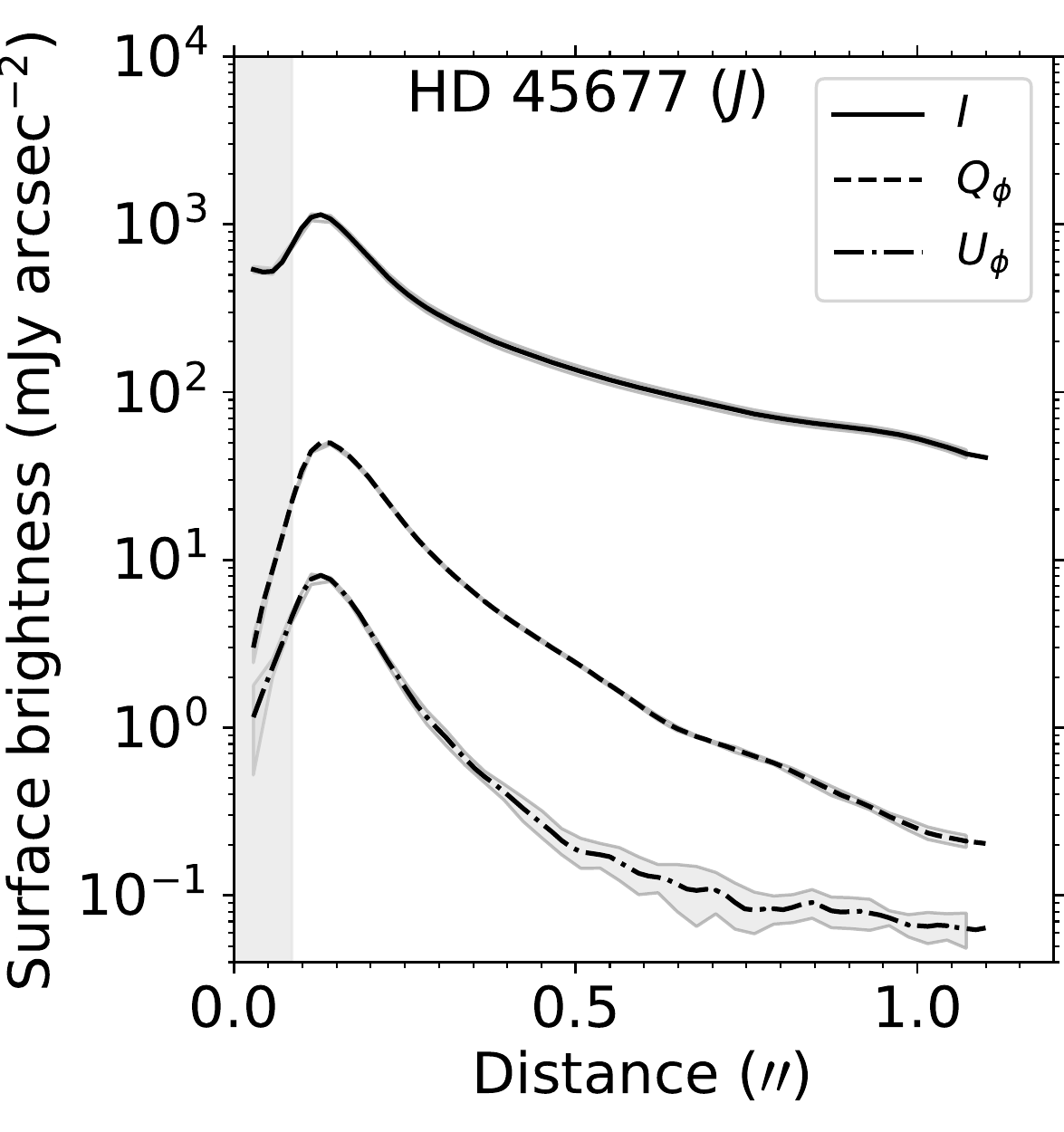}
	\includegraphics[width=0.23\textwidth, trim=0cm 0 0cm 0cm, clip]{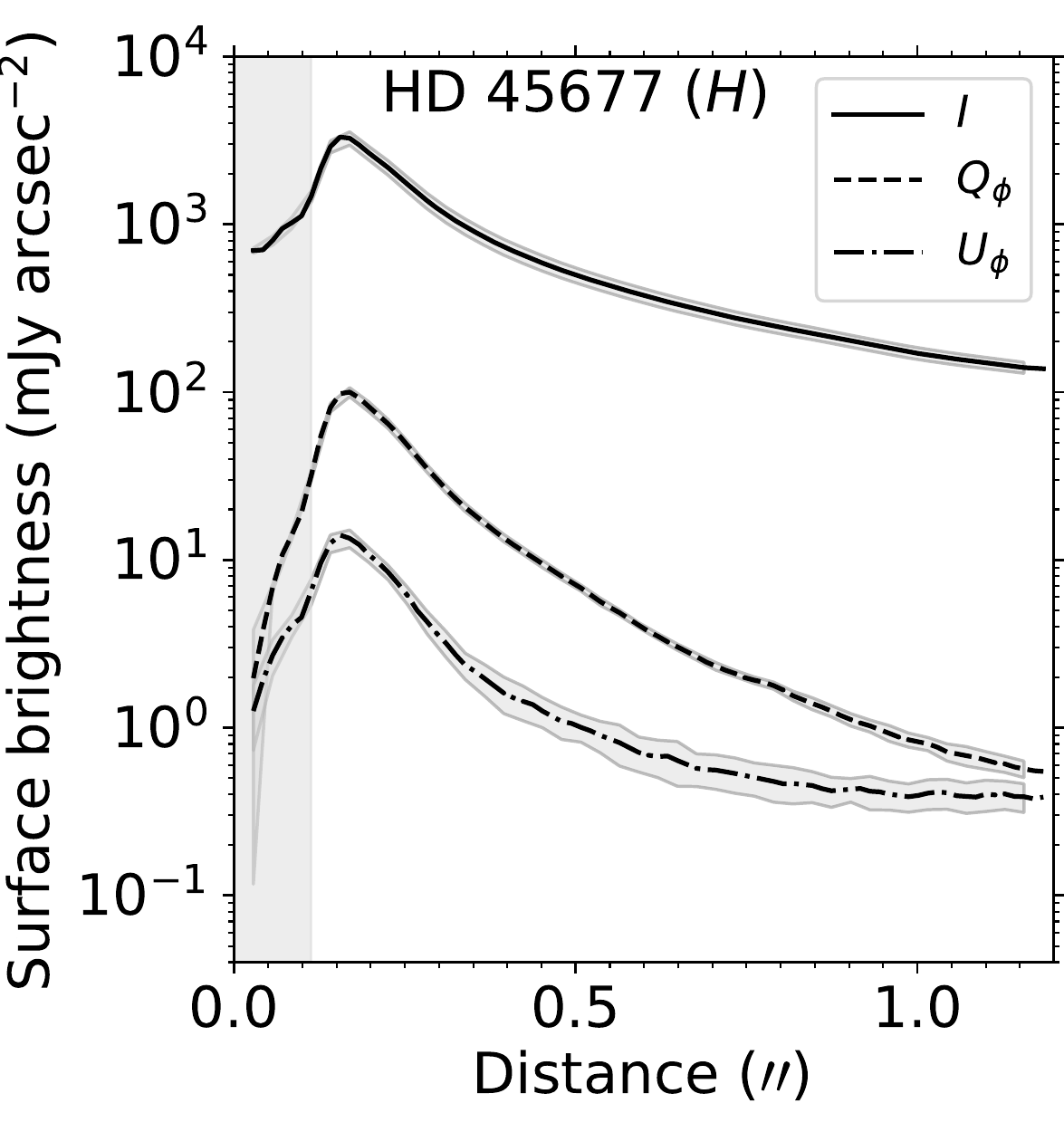}
	\includegraphics[width=0.23\textwidth, trim=0cm 0 0cm 0cm, clip]{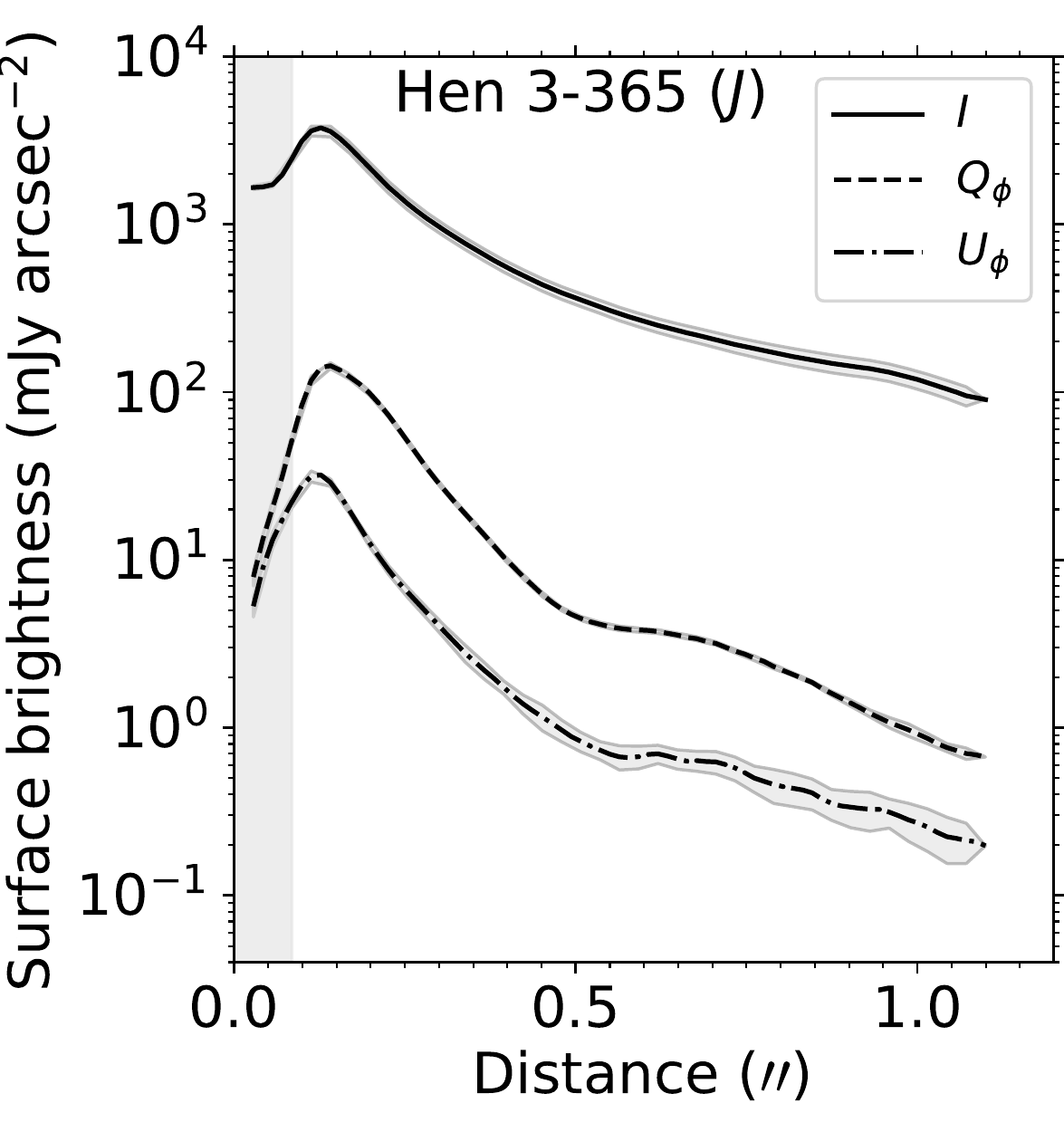}
    \caption{
    Surface brightness profiles in total intensity $I$ and radial Stokes maps $Q_\phi$ and $U_\phi$ for each of our observed datasets.
    We split the intensity maps into concentric annuli of width 1\,pixel
    and plot the annular radius against the mean of the absolute values within the annulus.  
    The shaded grey region around each profile shows the standard deviation of profiles across all interim stokesdc in each annulus, \edit{(as opposed to the standard deviation in each annulus)}.
    The vertical shaded region marks the extent of the coronagraphic spot.
    }
    \label{fig:radial_profiles}
\end{figure}

We present images of our five objects.
All objects are shown in total intensity ($I$) in Figure~\ref{fig:toti}
and in radial Stokes $Q_\phi$ and $U_\phi$ in Figures~\ref{fig:qphi}~and~\ref{fig:uphi} respectively.
Alternatively the objects are shown in linearly-polarized intensity in Appendix~\ref{app:lpi}.
In the following subsections we focus on each object individually, including the $Q_\phi$ and $U_\phi$ images in Figures~\ref{fig:qphi}~and~\ref{fig:uphi}.
Broadly speaking all of the $Q_\phi$ images show clear extended structures with irregular features.
First we cover the general features shared across total intensity and the radial profiles. 
 
We have applied a Gaussian smoothing with 2.1\,pixel standard deviation to all \textit{J}-band images in order to degrade the \textit{J}-band PSF to the same angular resolution as the \textit{H}-band PSF.
2.1\,pixel is equivalent to 30mas, the expected diffraction limit at \textit{J}-band for an 8\,m-class telescope.
In the total intensity images of Figure~\ref{fig:toti} we see a bright region immediately surrounding the coronagraphic spot. This is the stellar PSF and represents unpolarized light that will be absent from our final $Q_\phi$ and $U_\phi$ images.
The four satellite spots used for flux calibration are spaced evenly about the spot around halfway to the detector edge. 
In some images two spots are visible just outside the brightest part of the PSF (e.g. in FU~Ori, \edit{highlighted in Figure~\ref{fig:toti}}). 
These spots are not physically related to the central star
but result from a misalignment in the track assembly of GPI during the observations.
Light in these spots is unpolarized and does not impact our results in $Q_\phi$ and $U_\phi$, besides adding photon noise.

Companion candidates are revealed in the total intensity images for FU~Ori, MWC~789, and Hen~3-365. 
The companion candidate fluxes, separations and position angles are given in Table~\ref{tab:companions}.
We found the astrometry of the companion candidates by fitting a 2D Gaussian profile plus a sloping background to the Stokes \textit{I} image using a least squares fitting routine in python. Note that the companion candidate around FU~Ori is saturated, thus we masked out the saturated pixels and fit the Gaussian to the wings of the profile. 
\edit{Additionally, Hen~3-365~B (the inner most candidate) is located within the PSF wings of its primary star in the Stokes \textit{I} image, thus our companion fitting technique was not able to find a solution due to the strong slope of the primary's PSF. From the Stokes \textit{I} image, we created a low-pass image using a Gaussian kernel and then subtracted the low-pass image from the Stokes \textit{I} image to create a high-pass filtered image that retained the companion but removed the influence of the strong slope of the primary's PSF. We fit Hen~3-365~B within this high-pass filtered image and the resulting astrometry is reported in Table~\ref{tab:companions}.}
Finally, our FWHM measurements of the candidate companions Hen~3-365~B and MWC~789~B shown in Table \ref{tab:companions} are proximate of the spatial resolution of the images. 
The uncertainties in physical distance include the uncertainty associated with the distances in Table~\ref{tab:masses}.

Fluxes of companion candidates were estimated using aperture photometry in the total intensity \textit{I} images. 
The apertures were centered on the coordinates found using the Gaussian fit, and the aperture photometry was performed with the standard IDL prodecute "phot.pro".
Background annuli were selected to avoid regions where the PSF was pronounced.
\edit{
Companion candidates can appear with either positive or negative sign in the $Q_\phi$ and $U_\phi$ images, and this is only dependent on the objects' polarization position angle and not on any underlying physical parameter of the companion candidates.
The position angles are seen most clearly in the linearly-polarized intensity images in Appendix~\ref{app:lpi}.
}

Figure~\ref{fig:radial_profiles} shows radial profiles of the maps shown in Figures~\ref{fig:toti},~\ref{fig:qphi},~and~\ref{fig:uphi}.
We plotted the azimuthally averaged radial profile in I, \edit{$Q_\phi$, and $U_\phi$} maps with radial bins of 1 pixel in size.
As expected we see $Q_\phi$ profiles around two orders of magnitude fainter than the total unpolarized intensity profiles, and $U_\phi$ at least another order of magnitude fainter again. 
The errorbars for each target are calculated by creating a radial profile for each interim stokesdc. 
Then for each annulus the standard deviation across all profiles is used as the errorbar size in both positive and negative directions. 
\edit{In this way the errorbars are not a measure of the asymmetry in the discs, but a measure of random error across our stokesdc.}

The percentages of polarized flux are given in Table~\ref{tab:percpolflux}.
These are calculated by summing $Q_\phi$ in mJy in a large annulus spanning the coronagraph edge to the detector edge, and dividing by the 2MASS flux of the star.

\subsection{FU~Ori}

\begin{figure}
	\centering
	\includegraphics[width=0.47\textwidth]{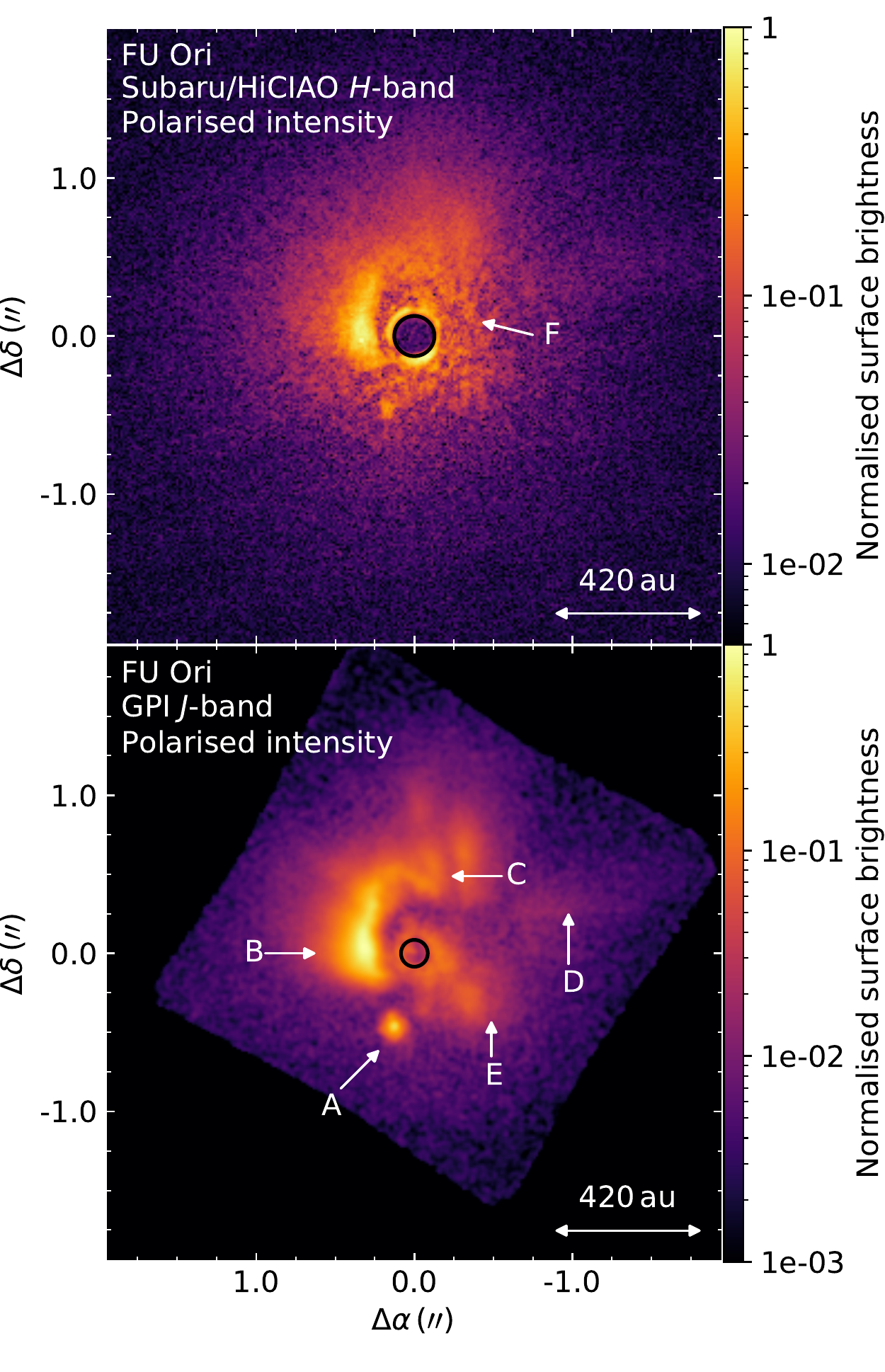}
    \caption{
      Linearly-polarized intensity images of FU~Ori.
      Top panel: The SUBARU/HiCIAO \textit{H}-band image \citep{2018ApJ...864...20T}.
      Bottom panel: The GPI \textit{J}-band image.
      The GPI image reveals the same structures at higher resolution.
      The following features are labelled: 
      \textbf{A}, the bright stellar companion, FU~Ori~S;
      \textbf{B}, a bright asymmetrical arm that extends around 200au towards the north and east directions;
      \textbf{C}, a dark stripe towards the northern tip of arm \textbf{A};
      \textbf{D}, a faint tail of material; 
      \textbf{E}, a diffuse region; and
      \textbf{F}, a feature identified in the HiCIAO image as a potential arm that does not appear in the GPI image.
      \edit{In both panels, the black circle marks the extent of the coronagraphic spot.}
    }
    \label{fig:fuori_lpi}
\end{figure}

FU~Ori shows a radically asymmetrical dust distribution. 
Figure~\ref{fig:fuori_lpi} highlights some of the features seen in $Q_\phi$, 
including the companion FU~Ori S (\textbf{A}), 
a bright arm that arcs from the east to the northern sides of the disk (\textbf{B})
before culminating in an infrared-dark stripe pointing north (\textbf{C}),
a faint tail of material stretching towards the west (\textbf{D}),
and a diffuse rounded region (\textbf{E}).
The bright arm (\textbf{B}) spreads around 200\,au (0.5\,'') north then 200\,au west and is bright enough to be faintly visible in the total intensity image (Figure~\ref{fig:toti}).
\textbf{B} resembles infalling streams of material observed around other objects with ALMA line emission (e.g. L1489 IRS, \citealp{2014ApJ...793....1Y}; HL Tau, \citealp{2019ApJ...880...69Y}).

The companion of FU~Ori (\textbf{A}), FU~Ori S, is located roughly 0.5'' (200au) from the central star at a position angle of $\sim 165^{\circ}$, in agreement with \citet{2004ApJ...601L..83W}~and~\citet{2004ApJ...608L..65R}.
FU~Ori~S appears with negative signal in $Q_\phi$, i.e. the region around the companion is radially polarized with respect to the central star. 
This polarization position angle is also seen in the Subaru-HiCIAO \textit{H}-band image \citep{2018ApJ...864...20T}.
The companion appears distinctively in the radial profile of Figure~\ref{fig:radial_profiles}, where there is an excess at 0.5'' in total intensity and in $U_\phi$ but not in $Q_\phi$.
It is possible that this companion is bright enough to cause the $U_\phi$ structures \edit{east of the spot in Figure~\ref{fig:uphi}} \edit{in the same area as} (\textbf{B}). 
In this case the $U_\phi$ would not indicate multiple scattering of light from the central star FU~Ori, but \edit{a} non-centrosymmetric polarization pattern due to light from FU~Ori~S.
This would explain why $U_\phi$ is so strong in the dusty region nearest FU~Ori~S, in the bright arm to the east of FU~Ori.
\edit{However we do not rule out the possibility that the $U_\phi$ structures are a result of multiple scattering from the central star.}

Figure~\ref{fig:fuori_lpi} also shows the Subaru-HiCIAO image of FU~Ori in scattered light from \citet{2018ApJ...864...20T}.
Generally we see the same features in both GPI and HiCIAO images,
though the GPI image more clearly shows the northern dark stripe (\textbf{C})
and the small round structures just south west of the spot  (\textbf{E}).
With GPI we recover the same faint western stream (\textbf{D}) but at a much reduced fraction of the maximum arm brightness.
We do not however recover structure \textbf{F} to the west of the spot in the Subaru-HiCIAO image, 
identified as a gas clump by \citet{2016SciA....2E0875L},
and instead see a gap between the stream and the dust immediately to the west of the spot.
We consider it unlikely that the non-detection of the HiCIAO feature \textbf{F} in our GPI image is due to temporal evolution in the density structure in the intervening 39 months or due to the different observing bands (H-band and J-band). Instead, we consider it more likely that feature \textbf{F} might represent residual Speckle noise in the HiCIAO image.
There is also the possibility that this temporal change is due to shadowing from the inner disk \edit{as in TW Hya \citep{2017ApJ...835..205D} and LKHa~330 \citep{2018AJ....156...63U}}, and the shadow is present in the GPI image but not the HiCIAO image.
The Speckle also appear in the same region as the faint feature \textbf{E}, which is indiscernible in the HiCIAO image.

There are just over three years between the Subaru-HiCIAO observation in October 2014 and the GPI observation in January 2018.
In this time there is no clear movement of FU~Ori~S with respect to the central star.
Cross-correlation analysis is unable to overlap the positions of the star in the two images by first translating and then rotating one image with respect to the other. 
Comparing separations and position angles indicates that the \edit{companion FU~Ori~S} has moved by a few 0.01\,'' and/or a few degrees, but this is less likely a physical movement of the companion than an inconsistency in the calculated central star position.
Having an incorrect central star position by even one pixel can cause changes to the separation and position angle on the order of the inconsistency between the Subaru-HiCIAO and GPI images. 
We calculate a back-of-the-envelope estimate for the orbital period $T$ of FU~Ori~S assuming that the observed separation $204\pm4$\,au is $r$, the radius of a circular orbit, and that the FU~Ori system has a mass  $1.5$\,$M_\sun$.
Using Kepler's third law ($T^2 \propto r^3$) we find $T\sim2400$\,yr.
Therefore in the three years between the HiCIAO and GPI observations the companion should have changed position angle on the order of $0.5$\,$^\circ$,
corresponding to a displacement of $\sim1.6$\,au or $4$\,mas - about 30\% of the size of one pixel in the GPI image.
Any change in companion position would be impossible to distinguish given that this value is on the order of our uncertainties.

In one theory of gravitational instability in disks, the growth timescale is at least as long as the orbital period at the outer disk edge \citep{1989ApJ...347..959A}.
This would lead to prohibitively long timescales for growth, and thus this scenario is unlikely to be responsible for the observed dust structures. However,
this picture ignores how disks are built up by infall from the protostellar envelope, and calculations including infall show that spiral structures can develop
in this case on shorter timescales \citep{2014ApJ...795...61B} or even gravitationally fragment into blobs \citep{2015ApJ...805..115V}.  Such models are more relevant as it is
generally thought that FU Ori outbursts occur during the protostellar infall phase \citep{1996ARA&A..34..207H}. In any case the timescales of outburst decay
show that the mass that is being accreted must come from inner disk regions regions $\lesssim 1$~AU
\citep{2007ApJ...669..483Z, 2010ApJ...713.1134Z}, well within our coronagraphic spot.

Given the similarity between the bright arm \textbf{B} and other infalling streams and the presence of the companion FU~Ori~S, it seems more likely that the vast majority of the observed structures are due to a complicated dusty environment rather than misaligned disks.

\begin{table*}
	\centering
	\caption{
    	Astrometry and fluxes of companion candidates or point sources present in the GPI images.
    }
	\label{tab:companions}
	\begin{tabular}{lllllll}
        \hline
        Object	&	Band	&	Flux   & Magnitude & Separation & Separation & PA \\
                &           &   (mJy)  & (mag)     & ('')       & (au)       & ($^\circ$)   \\
        \hline
        FU~Ori	        &	\textit{J}	
        & $ \phantom{00.0}$	--  
        & $ \phantom{00.0}$	--  
        & $ 0.489 \pm 0.020 $ 
        & $ \phantom{0}204 \pm \phantom{00}4 $ 
        & $ 164.2 \pm 0.3 $\\
        MWC~789         &   \textit{H}   
        & $ 26.5 \pm 0.1 $ 
        & $ 11.5 \pm 0.5 $
        & $ 0.426 \pm 0.018 $ 
        & $ \phantom{0}297 \pm \phantom{0}34 $ 
        & $ 216.3 \pm 0.3 $ \\
        Hen~3-365 (B)	&	\textit{J}	
        & $ \phantom{0}2.0 \pm 0.2 $
        & $ 14.8 \pm 3.1 $
        & $ 0.943 \pm 0.009 $ 
        & $ 1896 \pm 435 $ 
        & $ 101.2 \pm 0.1 $ \\
        Hen~3-365 (C)	&	\textit{J}	
        & $ 10.4 \pm 0.1 $
        & $ 11.9 \pm 0.4 $
        & $ 0.513 \pm 0.016 $ 
        & $ 1032 \pm 238 $ 
        & $ 133.1 \pm 0.2 $ \\
        \hline										
	\end{tabular}
	{\par Notes:
	The fluxes were measured in total intensity (Stokes \textit{I} in the stokesdc) using aperture photometry.
	Uncertainties in companion candidate fluxes use the same percentage uncertainty as for the flux scale factor presented in Table~\ref{tab:fluxcalib}.
	The distances and position angles with respect to the central star are also given. 
	Note that the flux of FU~Ori~S is not given as it is saturated.
	The separation in au incorporates the uncertainty in GAIA distance from Table~\ref{tab:background}. 
    }
\end{table*}

\begin{table}
	\centering
	\caption{
    The percentage of polarized flux in our images.
    }
	\label{tab:percpolflux}
	\begin{tabular}{lccc}
		\hline
		Object & \textit{J}-band (\%) & \textit{H}-band (\%) \\ 
		\hline
        FU~Ori    & 0.36\,$\pm$\,0.08 & --                   \\
        MWC~789   & --                & 0.60\,$\pm$\,0.09    \\
		HD~45677  & 0.72\,$\pm$\,0.16 & 0.90\,$\pm$\,0.14    \\
		Hen~3-365 & 0.80\,$\pm$\,0.17 & --                   \\
		HD~139614 & 0.27\,$\pm$\,0.06 & 0.35\,$\pm$\,0.06    \\
		\hline
	\end{tabular}
	{\par Notes:
    The uncertainties are calculated based on the dominating factor, the percentage uncertainty in the flux scale value (roughly 13\% for \textit{H}-band and 20\% for \textit{J}-band -- see Table~\ref{tab:fluxcalib}).
    }
\end{table}

\subsection{MWC~789}

We have discovered a stellar companion candidate and fine dust structures around MWC~789.
In the total intensity image (Figure~\ref{fig:toti}) we see a companion candidate and a disconnected arc to the north-west near the detector edge.
This arc is shown more clearly in the polarized $Q_\phi$ (Figure~\ref{fig:qphi}) - it is so extensive that it falls off the detector edge.
The bright central region is surrounded by small bright tails spreading outwards
and a trail of dust appears to link the companion candidate with the central region.
There is also an arm of material to the north-\edit{east} of the spot that falls off the detector edge.

Since we cannot see beyond the eastern side of the detector it is hard to say how the outer arc connects to the inner eastern arm.
The two arcs could meet in the east to form one large curved arm,
or alternatively there could be a transition disk-like structure as the outer arc shows an illuminated inner wall of an outer disk region with a dust clearing.
However the latter scenario would require an unusually large disk with radius $\gg$700au.
%
Both scenarios could be owing to the companion candidate,
either in generating the large arm or in causing the clearing.
The companion candidate could also be the cause of the smaller bright trails near the occulting spot,
particularly the trail that seems to connect the companion candidate to the central star.
\edit{Future analysis could include SED fitting with a large clearing to see whether this option is feasible.}

The companion candidate of MWC~789 appears at a distance of 0.426\,$\pm$\,0.018'' (297\,$\pm$\,34au) and a position angle of 216.3\,$\pm$\,0.3$^\circ$.
This companion candidate has not been reported in the literature previously.
Similarly to FU~Ori, the companion candidate is easily seen in the radial profile in Figure~\ref{fig:radial_profiles} as an excess at 0.4'' in total intensity and in $U_\phi$. 
This location also marks an increase in the sizes of the error bars for $Q_\phi$, which indicates that some surface brightness caused by the companion candidate appears to a lesser degree in $Q_\phi$. 

The \textit{H}-band polarization fraction of 0.60\,$\pm$\,0.09\% (Table~\ref{tab:percpolflux}) is on the same order as the 
\textit{I}-band polarization fraction of 0.95\,$\pm$\,0.48
and is consistent with the trend of falling fraction with increasing wavelength in visible light \citep{1995A&AS..111..399J}.

It seems unlikely that misaligned disks could be responsible for the complex strands near the companion candidate and coronagraphic spot. 
Therefore the circumstellar environment of MWC~789 likely has complex small-scale structures.

There are no previous high-resolution images of MWC~789 in the data archives of other extreme adaptive optics instruments or ALMA.
To constrain the dust distribution in the outer regions, MWC~789 should be observed with a larger field of view to discover how the northern arc is linked to the central star.

\subsection{HD~45677}
\label{sec:res_hd45677}

\begin{figure}
	\centering
	\includegraphics[width=0.45\textwidth]{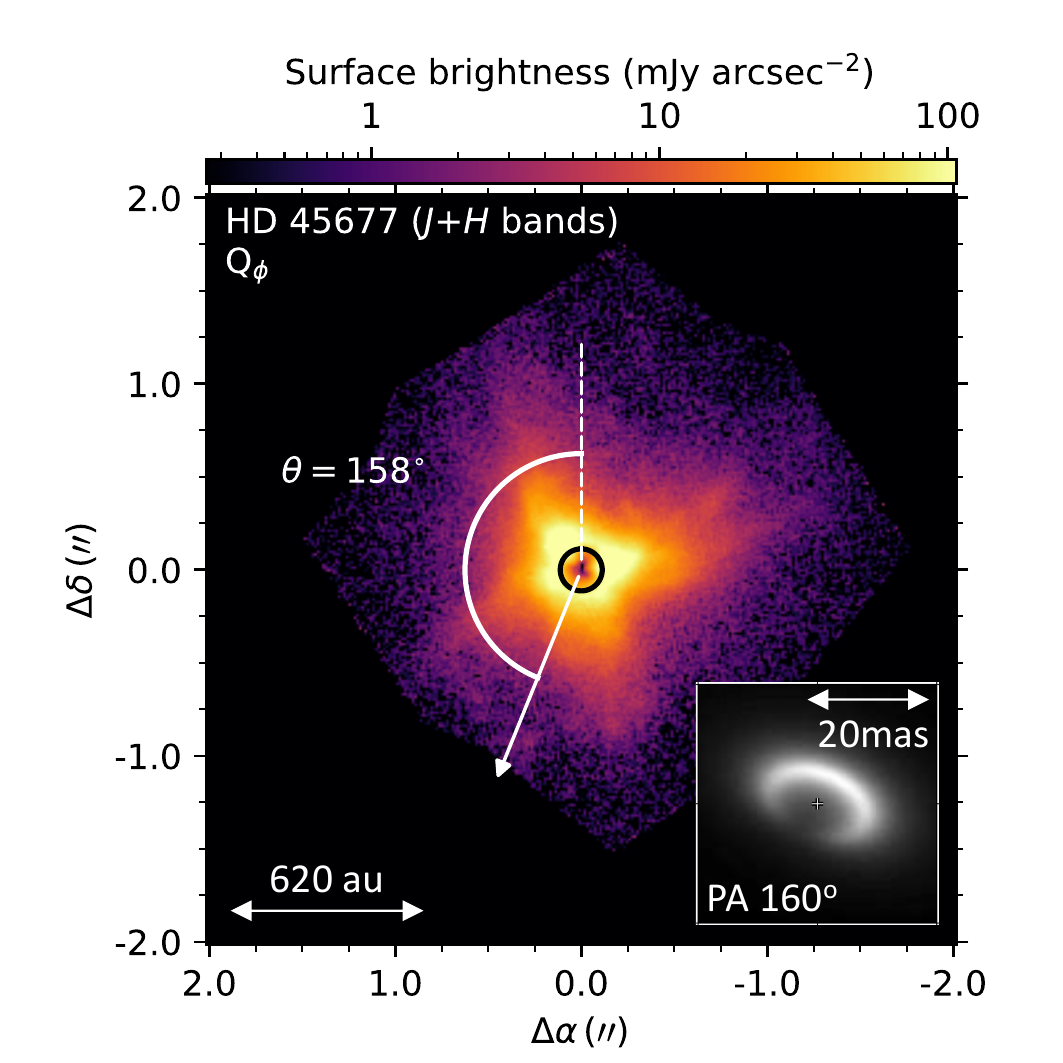}
    \caption{
    	Radial Stokes $Q_\phi$ image of HD~45677.
        \textit{J} and \textit{H} band images are combined linearly to improve the signal-to-noise ratio.
        Annotations show the symmetry position angle $\theta=158^\circ$ of the observed bipolar nebula structure as calculated in Section~\ref{sec:res_hd45677}.
        The inset image shows the model of the ring in the inner few milli-arcsec \citep{2006ApJ...647..444M} whose major axis shares a position angle with the larger scale emission.
    }
    \label{fig:hd45677_pa}
\end{figure}

\begin{figure}
	\centering
	\includegraphics[width=0.45\textwidth]{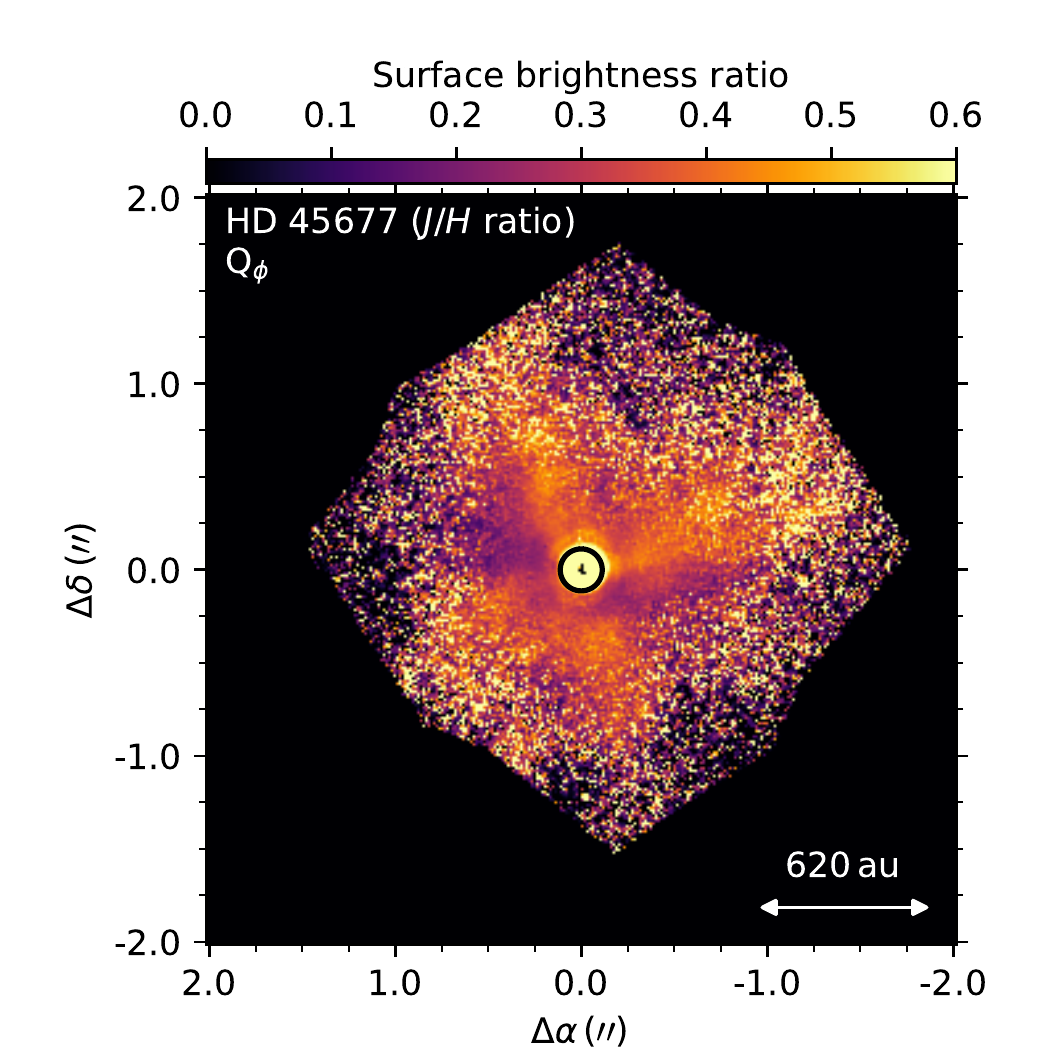}
    \caption{
        Radial Stokes Q$_\phi$ image of HD~45677 showing the \textit{J}-band image divided by the \textit{H}-band image. The two images were separately converted to mJy/arcsec$^2$ before combination, but were not normalised with respect to each other before division.
    }
    \label{fig:HD45677_j-h}
\end{figure}

The scattered light in HD~45677 reveals a symmetrical bipolar structure that resembles an edge-on view of an hourglass.
Aside from this large bipolar structure, we do not see the protoplanetary disk itself as it appears at far too small a size scale and is fully obscured by the coronagraph.
\edit{Our analysis here will focus on the broad shape and symmetry in the scattered light, and the interpretation of finer details will be discussed in a future paper.}
This object lacks any sharp-edged shadows that could result from misaligned disks.

\edit{moved these paragraphs from end of subsection:}
As noted above, the evolutionary state of HD~45677 is still rather uncertain. We have performed our own analysis using the UV photometry of \cite{1995yCat.2059....0T} and the $BVRI$ data compiled in \cite{2012AstL...38..331A}. We fitted Kurucz model atmospheres (with $\log g = 4$) to the photometry, using a grid search to determine the best fit stellar radius, reddening, and effective temperature whilst fixing the system's distance at 620\,pc (as in Table~\ref{tab:masses}). Our best fit has $T_{\rm eff} = 23\,000$\,K, $R=3.3$\,R$_\odot$, and $A_V = 0.6$. The fit is relatively poor, with a reduced chi-squared of 3, perhaps unsurprisingly given that the object displays photometric variability of over a magnitude on timescales of years and that the observations adopted were not contemporaneous.  
The best fit has a considerably higher $T_{\rm eff}$ than that determined by \cite{2018A&A...620A.128V}, $T_{\rm eff} = 16\,500^{+3000}_{-80}$\,K, but is  more in line with its early-B spectral classification. This luminosity and temperature are consistent with an 8\,M$_\odot$ mass star at an age of $10^6$\,years according to the PMS evolutionary models of \cite{2017ApJ...835...77M}, rather more massive and older than the \cite{2018A&A...620A.128V} estimate.

This analysis and the recent \citet{2018A&A...620A.128V} fits suggest that HD~45677 is a pre-main-sequence object.
In this case the observed bipolar emission structure would be a protostellar envelope.
This structure is common in objects in star-forming regions 
\citep{1991ApJ...378..611T}.
However the emission nebula does not rule out HD~45677 being a more evolved object, e.g. a planetary nebula. 
\edit{Some bipolar planetary nebulae bear strong resemblance to the general shape of HD~45677 (\citealp[e.g. the ``Squid Nebula'',][]{2012RMxAA..48..223A}; \citealp[``Hubble 12'',][]{2007ApJ...660..341K}).
}
There is still the evidence for HD~45677 having a bipolar outflow (see Section~\ref{sec:ourtargets}), in which case the observed emission nebula structure could mark the walls of an outflow cavity.
The evolutionary status of this object is still ambiguous and makes HD~45677 a prime candidate for follow-up modeling \edit{and follow-up observations}.
\edit{end of moved content}

Figure~\ref{fig:hd45677_pa} shows the combined \textit{J}+\textit{H} image with the position angle of the hourglass structure, which we find by assuming that the underlying structure is axis-symmetric.
The right-hand side of the image is flipped horizontally and laid over the left-hand side, then the flipped half is subtracted from the original values. 
A symmetrical image gives weaker residual structure remaining in the image after subtraction. 
The strength of the residual structure is measured with the sum of the absolute value of the residuals, 
with the sum using only the central half of the image to reduce the effect of values near the detector edge.
The original image is then rotated and the residual sum calculated again in the same way.
We use the built-in scipy.optimize.minimize function with the ``Nelder-Mead" method to minimise the residual sum and so find the rotation angle that produces the most symmetry. 
Using this method with the combined J+H $Q_\phi$ image, we determine the symmetry axis position angle to 158$^{\circ}$.
For completeness the values determined from the \textit{J}-band and \textit{H}-band images separately are $157^\circ$ and $158^\circ$ respectively.

Assuming stronger forward scattering, the brighter north-western half of the emission nebula shape must be inclined towards the observer.
Alternatively the difference in brightness could be owing to an inner disk structure obstructing light from the farther lobe \citep{1991ApJ...378..611T}.
The second scenario would match the inner ring model from \citet{2006ApJ...647..444M}, which is highlighted in Figure~\ref{fig:hd45677_pa} 
and has a similar minor axis position angle to the angle of symmetry in the hourglass structure - 160$^\circ$ and 158$^\circ$ respectively.
Both position angles are comparable with the polarization and blowout angles of $164^\circ$ and $175^\circ$ respectively \citep{2006MNRAS.373.1641P},
which establishes a connection between the inner ring and the larger-scale structure.

Figure~\ref{fig:HD45677_j-h} shows the ratio of the \textit{J} and \textit{H} $Q_\phi$ images.
The emission is \edit{somewhat} redder along the major axis of the unseen disk and bluer in the region of the bipolar structure.
\edit{These features are clearer nearest the center of the image, and the underlying structure of the bipolar nebula is lost to the background scatter at larger radii.}
\edit{Near the center, the difference in surface brightness is distinctive and following the bipolar shape enough to indicate that there is some systematic difference between the scattering in \textit{J} and \textit{H}-bands.}
The reddening could be caused by extinction or different grain sizes in the different regions.
\edit{Separately,} both wavelengths reveal the same finer structures such as distinct wisps or streams leading out of the emission nebula shape.

\subsection{Hen~3-365}

Our results show the first resolved high-contrast imaging of Hen~3-365.
Hen~3-365 shows another remarkably asymmetric circumstellar environment. 
Immediately west of the occulting spot we detect a rounded bright region, 
and there are additional arcs of material that are disconnected from this brighter region. 
The brightest, in the north-west, lies in a straight line $\sim$2000\,au long at an angle almost tangential to the connecting line to the central star.
This region also is distinguishable in total intensity and the $U_\phi$ image, which suggests there could be multiple scattering effects. 
The structure appears least significantly in the radial profile of total intensity (see Figure~\ref{fig:radial_profiles})
but is clearly visible in both $Q_\phi$ and $U_\phi$ profiles.
There are no obvious structures that correspond to the EW and N outflows as in \citet{2006MNRAS.367..737B}.

In addition there is a dark region immediately to the east of the coronagraphic spot. 
This region is negative in $Q_\phi$ and is also distinguishable in $U_\phi$.
It appears to be physical in origin.
The effect could have been exacerbated during the reduction process due to butterfly patterns from a sub-par stellar polarization removal (see Appendix~\ref{app:reduction_methods}), but this dark region does appear to some extent regardless of reduction method.
The physical size of this dark region is so large that it is unlikely to be caused by a misaligned disk structure that would be much smaller in scale.
The dark region could be an imprint of the companion candidates
if they have cleared this region,
since the three features appear in the same range of position angle.

Hen~3-365 shows two faint point sources that we designate B and C.
Their fluxes, separations and position angles are given in Table~\ref{tab:companions}. 
The source Hen~3-365-B is considerably fainter than Hen~3-365-C and the companion candidate star around MWC~789, at only $\sim$10--20\,\% of the latter objects' fluxes. 
The two point sources appear with negative $U_\phi$ 
and are located away from the central stars, with distances on the order 1000\,au.
Because the point sources are so small and faint it is unlikely that they have distorted the $Q_\phi,U_\phi$ images to create false evidence of multiple scattering.
In particular they are far away in physical and projected distance from the north-western arc that shows structure in $U_\phi$ and are unlikely to be responsible for this structure.
Any association of the point sources and the central star should be investigated with follow-up observations to see if the point sources are co-moving.

\subsection{HD~139614}

\begin{figure}
	\centering
	\includegraphics[width=0.45\textwidth]{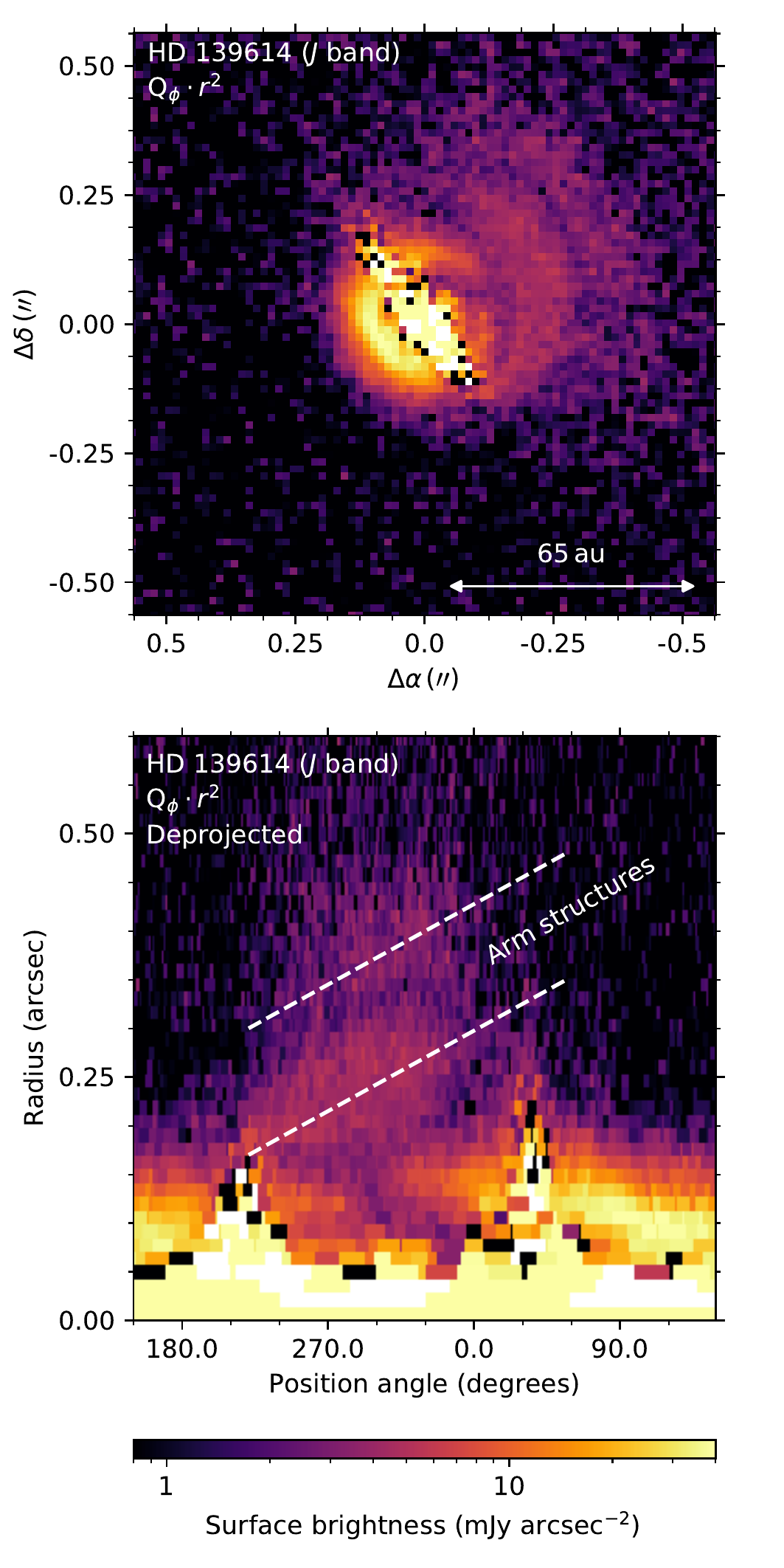}
    \caption{
        Spiral arms in HD~139614 ($Q_\phi$, $J$-band, \edit{normalized by distance from the image center $r$ by multiplying by $r^2$}).
        These data were taken without the use of a coronagraph, 
        resulting in saturation in the image center.
        The saturated area appears as a stripe centered at (0.0,0.0$''$) of position angle $\sim30^\circ$ and extent $\sim0.3''$
        where some of the affected values are colored black.
        The below panel shows the same image deprojected into polar coordinates.
        The marked arm structures increase in radius and are likely spiral in nature.
        }
    \label{fig:HD139614_unsharp}
\end{figure}

For HD~139614 \textit{J}-band observations, the 2019 data were used in all cases except for examining arm structures in Figure~\ref{fig:HD139614_unsharp}, where the 2017 data without the coronagraphic spot \edit{are} presented using the 2017 data with the coronagraph for flux calibration.

HD~139614 shows faint emission in $Q_\phi$ out to $\sim$150au in all directions around the object (particularly on the western side)
and a brighter region within 50au of the center, consistent with predictions that the outer disk extends to 75--100au \citep{2006MNRAS.373.1641P,2014SPIE.9146E..2TL}.
Depending on the positioning of the coronagraphic spot, the structure at 50\,au could correspond either to a ring or a tight spiral arm. Our three independent observations show \edit{that this structure lies} consistently to the north of the star, which strongly supports the tight spiral arm interpretation.
The spiral arm originates from the north and continues anticlockwise around the image.
Its inner edge is just visible in data taken without coronagraphy shown in Figure~\ref{fig:HD139614_unsharp}.
It is impossible to flux calibrate an image without coronagraphy using the DRP methods described in Section~\ref{sec:reduction}, 
so we use the flux scale from contemporary data taken with the coronagraph.
Both stars in the HD~139614 binary system appear saturated in the image without the coronagraph and cause a distorted band near the center of the image.
It is not possible to locate the two stars behind this distortion.

Multiple faint arms to the northwest are easier to see in Figure~\ref{fig:HD139614_unsharp} 
%
where we present a deprojected version of the $Q_\phi$ image that is translated into polar coordinates. 
\edit{To make this image, we created} a list of coordinates that are regularly spaced in ($r$,$\phi$) space -
every 1.0 pixel or 0.0141$''$ in radius from the image center to detector edge, and every 1 degree in position angle $\phi$ 
- and sample the values in these coordinates from the Cartesian image.
\edit{In addition the images in this figure are distance-normalized, i.e. the surface brightness values are scaled by the distance from the image center $r$ in arcsec, to more clearly show the fainter outer arms.
}

The deprojected image shows that the arm structures generally increase in distance from the image center.
This is most consistent with the arms being spiral structures. 
\edit{The outermost arm begins to decrease in radius again at a position angle of 300$^\circ$, which makes it more likely that it is instead part of an elliptical ring or other irregular feature.}
The apparent gap between the two arm structures could be due to a shadow cast by the inner arm \edit{onto the outer arm}.
There is a break in the bright innermost arm (radius 0.13'', PA 300$^\circ$ in Figure~\ref{fig:HD139614_unsharp}) that is likely due to a shadow cast by a misaligned disk.
This break appears in all of our HD~139614 datasets \edit{and so is not due to any observing effects such as saturation of the central star or the affected region being covered by the coronagraph}. 
Notably there is little structure in $U_\phi$ (Figure~\ref{fig:uphi}) for this object compared with the other objects presented here, 
which suggests that multiple scattering is negligible in the resolved  \edit{arm structures}.

\edit{The two highlighted arm structures are too faint to be seen outside a certain range of position angle $\sim$210--0$^\circ$, i.e.}
the polarized emission is brighter on the western side. 
\edit{In the simple first approximation, this brightness asymmetry is} consistent with a strong forward-scattering regime and the western side inclined towards us.
This matches an inner disk model with an inclination 20$^\circ$ and position angle 112$^\circ$ as fitted to MIDI visibilities \citep{2014A&A...561A..26M}. 
The VLT/CRIRES prediction of a close-in planet could result in a skewed, tilted inner disk \citep{2017A&A...598A.118C}, leading to the observed asymmetry in scattered light.
\edit{A more robust estimate of the inclination of the object and its scattering phase function will be left to future analysis.}

The inner disk and gap predicted by the literature lie under the occulting spot in our images.
The Subaru-HiCIAO scattered light non-detection \citep{2010PASJ...62..347F} used a coronagraph with 1.0'' spot diameter coronagraph, which is much larger than the order 0.1'' diameter spot used in our GPI observations.
As such it is likely that even the fainter regions we detect are covered by the coronagraph in the Subaru-HiCIAO observation.

\section{Conclusions}
\label{sec:conclusions}

We have observed five objects with extended irregular dust structures in their circumstellar environments
in high-resolution
using GPI in PDI mode with extreme adaptive optics and coronagraphy. 
Though we present all five as young objects in the protoplanetary disk phase, it is unclear which structural formation process is dominant.
The features could be explained either as the complex environment around 
extremely young disks with complex initial conditions
, or alternatively as shadows and uneven illumination from misaligned disks.
Some of our objects show companions or companion candidates
which could also be responsible for the messy structures,
for example the clearing around MWC~789
and the negative-$Q_\phi$ region in Hen~3-365.
The objects tend to have strong structures in $U_\phi$ indicating a complex environment that could be due to multiple scattering or illumination from the stellar companion candidates.

Each individual object shows interesting features.
FU~Ori has a complex highly irregular morphology
with a companion and bright arm. 
We resolve more structure than the existing Subaru-HiCIAO image 
including a dark stripe to the north
Any apparent changes from the Subaru-HiCIAO image are likely to be due to reduction in Speckles,
and we estimate that any movement of the companion would be too slight to observe.

We made the first high-contrast detection of MWC~789, detecting a previously unknown companion candidate and an extended and apparently disconnected arc.
There is potentially a wide clearing between the central star and this distant arc, however this should be confirmed with further observations with a wider field of view.
This clearing could be due to the presence of the companion candidate.

The evolutionary status of HD~45677 has been contested for some time. 
Our observations show a bipolar emission structure
and our preliminary analysis
indicates that it is likely a young object with 
an extended dusty circumstellar environment.
The symmetry position angle of the nebula 
is comparable with the minor axis position angle of a ring model fit to \textit{H}-band IOTA interferometeric visibilities,
establishing a link between the small- and large-scale structures.

We present the first high-resolution image of Hen~3-365 
in which 
we newly identify two companion candidates,
along with multiple arc structures.
Assuming a correct GAIA distance of 2000\,pc, the physical scales of these arcs span thousands of au.

HD~139614 has a tight spiral structure and asymmetrical brightness.
The innermost spiral arm contains a break that could be due to shadowing from a misaligned disk.

Further observations of these objects in other wavebands
could be combined with radiative transfer modeling
to find out more about the hidden structures of the dust in these circumstellar environments.
A particularly good target for follow-up observations is MWC~789, where a larger field of view than GPI would reveal the connection between the central star and distant arc.
Follow-up observations of Hen~3-365 could explain the nature of the two nearby point sources and whether they are co-moving with the main system.

\section*{Acknowledgements}

The authors wish to thank Fredrik Rantakyro, whose help with various instrumentation difficulties enabled us to finish our observations.
We also wish to thank Bruce MacIntosh for kindly allowing us use of his remote observatory.
Finally we wish to thank Michihiro Takami for sharing his reduced HiCIAO data for FU~Ori and allowing us to present it here.

A.L. wishes to thank the Science Technology and Facilities Council (STFC) for the studentship that supported this work (project reference 1918673).
JM and ER acknowledge funding from a National Science Foundation grant (reference NSF-AST1830728).
S.K. acknowledges support from an European Research Council Starting Grant (Grant Agreement No.\ 639889).

This paper is based on observations obtained at the Gemini Observatory, which is operated by the Association of Universities for Research in Astronomy, Inc., under a cooperative agreement with the NSF on behalf of the Gemini partnership: the National Science Foundation (United States), the National Research Council (Canada), CONICYT (Chile), Ministerio de Ciencia, Tecnolog\'{i}a e Innovaci\'{o}n Productiva (Argentina), and Minist\'{e}rio da Ci\^{e}ncia, Tecnologia e Inova\c{c}\~{a}o (Brazil).

This work has made use of data from the European Space Agency (ESA) mission
{\it Gaia} (\url{https://www.cosmos.esa.int/gaia}), processed by the {\it Gaia}
Data Processing and Analysis Consortium (DPAC,
\url{https://www.cosmos.esa.int/web/gaia/dpac/consortium}). Funding for the DPAC
has been provided by national institutions, in particular the institutions
participating in the {\it Gaia} Multilateral Agreement. 
This publication makes use of data products from the Two Micron All Sky Survey, which is a joint project of the University of Massachusetts and the Infrared Processing and Analysis Center/California Institute of Technology, funded by the National Aeronautics and Space Administration and the National Science Foundation.
%




\bibliography{gpi_data_paper} 



\appendix

\section{Data reduction - stellar polarization subtraction}
\label{app:reduction_methods}

\begin{figure}
	\centering
	\includegraphics[width=\textwidth]{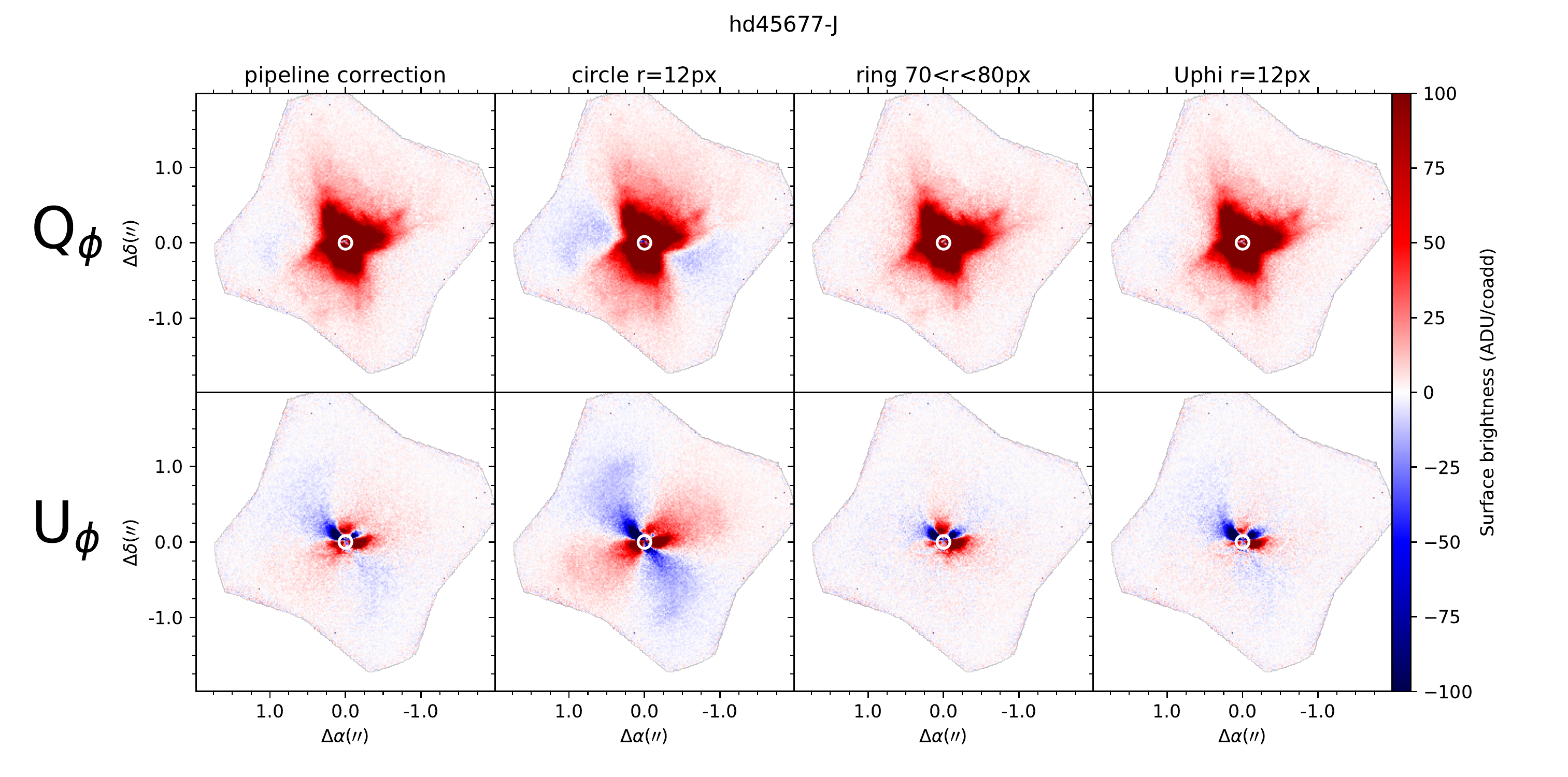}
    \caption{
    	Comparison between $Q_\phi$ (top row) and $U_\phi$ (bottom row) images from the various stellar polarization subtraction methods \edit{described in Appendix~\ref{app:reduction_methods}} for HD~45677 in \textit{J}-band.
    	The linear color scale is truncated to $\pm100$ counts to highlight the smaller changes between methods.
    }
    \label{fig:reduction_methods}
\end{figure}

Unlike many other objects in the literature, the presented objects have strong $U_\phi$ structures. 
This appendix covers the methods used to investigate whether the structures were physical or were introducted during the data reduction process.

Initially we investigated whether some of the structure appeared due to an incorrect stellar polarization removal. 
As well as modifying the sign of the observed structures, this can introduce a ``background" signal in $U_\phi$ as seen in Figure~\ref{fig:reduction_methods}.
Due to the way that $U_\phi$ is created (see Equations~\ref{eqn:qphi}~\&~\ref{eqn:uphi}), adding a factor of total intensity $I$ to Stokes maps $Q$ and $U$ tends to move the background farther from zero. 
This appeared in our images as the map being split into four squares - a butterfly pattern - with the two squares in opposite corners having positive signal and the other two corners showing negative signal (in Figure~\ref{fig:reduction_methods} this appears as corners appearing strongly red or blue). 

The method employed by the DRP \edit{(``pipeline correction'' in Figure~\ref{fig:reduction_methods})} involves subtracting a small fraction of total intensity from the other Stokes cubes.
The exact scaling factor is determined by considering the region covered by the coronagraphic spot where no disk-related scattered light is expected \edit{(radius r$<$8pix for \textit{J} and r$<$6pix for \textit{H})}. 
In this region we define two scaling factors:
\begin{equation}
    f_Q = \frac{\textnormal{mean(}Q\textnormal{)}}{\textnormal{mean(}I\textnormal{)}}
\end{equation}
\begin{equation}
    f_U = \frac{\textnormal{mean(}U\textnormal{)}}{\textnormal{mean(}I\textnormal{)}}    
\end{equation}
These are used to define a new Q and U across the full maps, $Q^\star$ and $U^\star$:
\begin{equation}
    Q^\star = Q - I\times f_Q
\end{equation}
\begin{equation}
    U^\star = U - I\times f_U
\end{equation}
This $Q^\star$ and $U^\star$ are used to create $Q_\phi$ and $U_\phi$ as in Equations~\ref{eqn:qphi}~\&~\ref{eqn:uphi}.
Using the default reduction recipes from the GPI DRP did not successfully remove the background pattern in all cases, so another method was required.

We considered a similar $U_\phi$ minimisation method to that presented by \citet{2018ApJ...863...44A} \edit{(``Uphi r=12px'' in Figure~\ref{fig:reduction_methods})}. 
For objects with no multiple scattering the radial Stokes map $U_\phi$ should contain little to no physical signal. 
Any observed signal is likely owing to instrumental polarization effects, which can be countered by minimising the signal in $U\phi$, i.e. minimise $\textnormal{sum}(\textnormal{abs}(U_\phi))$ for all values within an 85pixel radius of the center (spanning outwards as close to the detector edge as possible).
For the correction, the Stokes maps were weighted according to these parameters: $\gamma$, c$_1$, c$_2$, c$_3$, c$_4$. 
The weighted Stokes maps $Q^{\star}$ and $U^{\star}$ had the following form:
\begin{equation}
Q^{\star} = Q + c_1 I + c_2
\end{equation}
\begin{equation}
U^{\star} = U + c_3 I + c_4
\end{equation}
An optimisation function, scipy.minimize from python's scipy module, found the values of $\gamma$, c$_1$, c$_2$, c$_3$, c$_4$ that minimise sum(abs($U_\phi$)).
Then the corrected stokesdc containing the images $[I,Q^\star,U^\star,V]$ is used to create $Q_\phi$ and $U_\phi$ as in Equations~\ref{eqn:qphi}~\&~\ref{eqn:uphi}.
This method has successfully revealed faint structures in other objects in our survey (not presented here).
However the presented objects as well as HD~34700 \citep{2019ApJ...872..122M} tend to show some structure in $U_\phi$ that could have physical causes, such as multiple scattering in HD~34700. 
For this reason these objects should not be reduced using a method designed to squash all signal in $U_\phi$ to zero, as this could be removing evidence of actual astrophysical effects.

For this reason we created an improved version of the DRP stellar polarization subtraction employing the same $f_Q, f_U$ subtraction technique over areas of different sizes.
For these tests we used a separate circular region with radius $r\leq12$pix and  annular region covering $70\leq r \leq80$pix \edit{(respectively ``circle r=12px'' and ``ring 70$<$r$<$80px'' in Figure~\ref{fig:reduction_methods})}, compared with the coronagraphic spot size of 6 or 8 pixels for \textit{J} and \textit{H}-band images respectively. 
The circle was chosen since ordinarily there is reduced useful signal close to the spot due to the stellar PSF, and the ring was chosen to cover more of the background signal.
For our more extended objects there was a possibility of physical structures appearing even as far out as the ring.
To test whether this would be an issue we searched for contamination in these regions - for pixels where $\frac{Q_\phi}{I}>0.1$. 
We found that this condition was not met in the ring for any objects, but was barely met in the circle for Hen~3-365 for a few pixels just outside the spot with values no larger than 0.112.

A comparison of the four methods for one object, HD~45677 in \textit{J}, is shown in Figure~\ref{fig:reduction_methods}. 
For most of the presented objects the reduction made no difference to the appearance of $Q_\phi$, however for HD~45677 there is a strong negative signal introduced to the east and west when using the circle reduction method. 
Given that the reduction method can affect even structures in $Q_\phi$ it is especially important to use the optimal method. 
The differences between methods are even more pronounced in $U_\phi$. The DRP and circle methods give the strong square background pattern, and this effect is visible to a much reduced extent using the $U_\phi$ minimisation. 
The effect is almost entirely removed with the ring method.
We find that this holds for all of our presented objects - the ring method consistently removes the background square pattern much better than the other methods.

A comparison of the calculated $f_Q$,$f_U$ across all data cubes for each object finds that the values tend to be equally consistent between the circle and ring methods.
For some objects we find outliers, e.g. the ring method in MWC~789 calculates inconsistent $f_Q$,$f_U$ in two cycles but the circle method gives consistent $f_Q$,$f_U$ for all cycles.
There does not appear to be a significant test of how the consistency of $f_Q$,$f_U$ leads to a ``better" reduction.

Two of the presented objects have been observed in two wavebands, \textit{J} and \textit{H}.
The orientation of the background squares pattern is consistent between wavebands, but can vary between the stellar subtraction methods. 
The resulting $Q_\phi$, $U_\phi$ images are broadly consistent between wavebands for each method though the strength of the square pattern can vary.

Due to its superior removal of the stellar polarization, the ring method was used to reduce the objects presented throughout this paper.

\section{LPI}
\label{app:lpi}
\begin{figure*}
	\centering
    \includegraphics[width=0.8\textwidth]{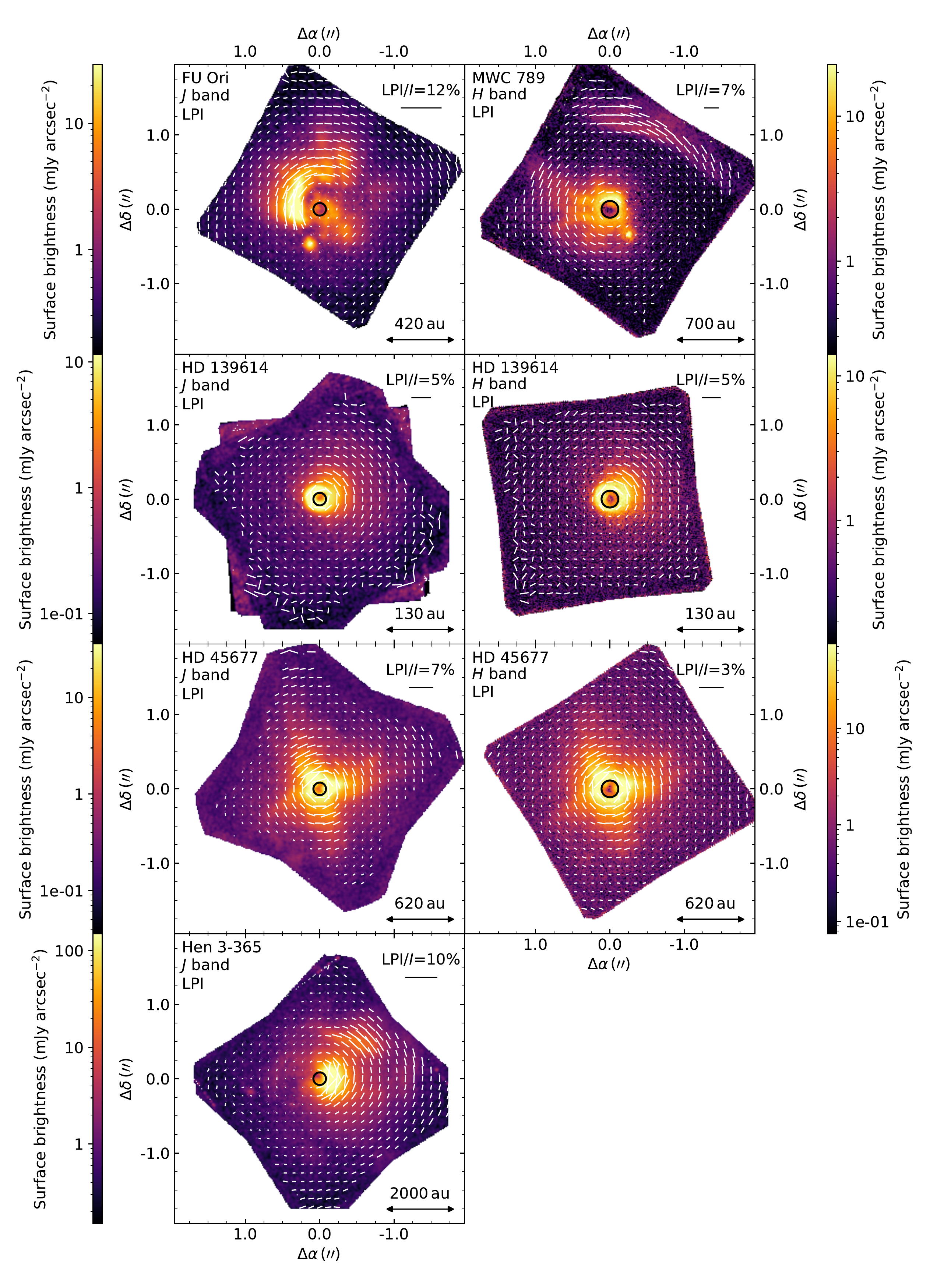}
    \caption{
    	Linearly-polarized intensity (LPI) images for our observed targets with polarization vectors overlaid. 
    	North is up, East is left.
    	The physical scales are the same as in Figure~\ref{fig:toti}.
    }
    \label{fig:lpi}
\end{figure*}

Figure~\ref{fig:lpi} shows our objects in linearly-polarized intensity with polarization vectors to show the position angle across the image.
The Figure is provided as an alternative to the $Q_\phi,U_\phi$ representation in Figures~\ref{fig:qphi}~and~\ref{fig:uphi} respectively.
Broadly speaking the images show the same structures as $Q_\phi$ with the addition of any radially-polarized regions that are negative in $Q_\phi$ and so are absent from Figure~\ref{fig:qphi},
for example the companion FU~Ori~S and the dark region to the east of the spot in Hen~3-365.

\section{ORCID iDs}

Anna Laws \url{https://orcid.org/0000-0002-2145-0487}

Tim J. Harries \url{https:/orcid.org/0000-0001-8228-9503} 

Benjamin R. Setterholm \url{https:/orcid.org/0000-0001-5980-0246} 

John D. Monnier \url{https:/orcid.org/0000-0002-3380-3307}  

Evan Rich \url{https://orcid.org/0000-0002-1779-8181}

Jaehan Bae \url{https:/orcid.org/0000-0001-7258-770X}  

Alicia Aarnio \url{https://orcid.org/0000-0002-1327-9659}

Fred C. Adams \url{https://orcid.org/0000-0002-8167-1767} 

Sean Andrews \url{https://orcid.org/0000-0003-2253-2270} 

Nuria Calvet \url{https://orcid.org/0000-0002-3950-5386}

Catherine Espaillat \url{https:/orcid.org/0000-0001-9227-5949}  

Lee Hartmann \url{https:/orcid.org/0000-0003-1430-8519}  

Sasha Hinkley \url{https://orcid.org/0000-0001-8074-2562}

Andrea Isella \url{https://orcid.org/0000-0001-8061-2207}

Stefan Kraus \url{https:/orcid.org/0000-0001-6017-8773}  

David Wilner \url{https:/orcid.org/0000-0003-1526-7587}  

Zhaohuan Zhu \url{https:/orcid.org/0000-0003-3616-6822}


\end{document}